\documentclass[draft]{ecai}  

\usepackage{graphicx}
\usepackage{latexsym}
\usepackage[numbers]{natbib}

\paperid{1742}        

\usepackage{paralist}

\usepackage{csquotes}

\usepackage{thmtools,thm-restate} 

\usepackage{algorithm}
\usepackage[noend]{algorithmic}

\usepackage{xargs}

\usepackage{xspace}

\usepackage{xstring}

\usepackage{boolexpr}

\usepackage[cmex10]{mathtools}
\usepackage{amsfonts}
\usepackage{mathrsfs}
\usepackage{textcomp}
\usepackage{pifont}
\usepackage{amssymb,bm,xargs}
\usepackage[shortlabels]{enumitem}

\usepackage{wrapfig}

\usepackage{lipsum}
\usepackage[obeyDraft]{todonotes}
\usepackage{cancel}

\newtheorem{definition}{Definition}
\newtheorem{lemma}{Lemma}

\newenvironment{proof}{\emph{Proof.} }{\hfill $\square$ \\}

\newcommand{\argemp}[2]{\if&#1&\else#2\fi}

\newcommand{\argdef}[2]{\if&#1&#2\else#1\fi}

\newcommand{\argint}[3]{\if&#2&\else#1#2#3\fi}

\newcommand{\argext}[3]{\if&#1&#3\else#1\if&#3&\else#2#3\fi\fi}

\newcommandx{\mthfnt}[3][1=, 2=0]{{
	\IfStrEqCase{#1}
	{%
		{}%
		{#3}%
		{Name}%
		{%
			\IfStrEqCase{#2}
			{%
				{0}{\mathcal{#3}}%
				{1}{\mathscr{#3}}%
				{2}{\mathfrak{#3}}%
				{3}{\mathbb{#3}}%
			}
			[\ensuremath{\clubsuit}]%
		}%
		{Set}%
		{%
			\IfStrEqCase{#2}
			{%
				{0}{\mathrm{#3}}%
				{1}{\mathsf{#3}}%
				{2}{\mathbb{#3}}%
				{3}{\mathbf{#3}}%
			}
			[\ensuremath{\clubsuit}]%
		}%
		{Fun}%
		{%
			\IfStrEqCase{#2}
			{%
				{0}{\mathsf{#3}}%
				{1}{\mathrm{#3}}%
			}
			[\ensuremath{\clubsuit}]%
		}%
		{Rel}%
		{%
			\IfStrEqCase{#2}
			{%
				{0}{\mathit{#3}}%
				{1}{\mathtt{#3}}%
			}
			[\ensuremath{\clubsuit}]%
		}%
		{Sym}%
		{%
			\IfStrEqCase{#2}
			{%
				{0}{\mathtt{#3}}%
				{1}{\mathbf{#3}}%
			}
			[\ensuremath{\clubsuit}]%
		}%
		{Elm}%
		{\mathnormal{#3}}
	}
[\ensuremath{\clubsuit}]%
}}

\newcommand{\mthsub}[1]{\argemp{#1}{\ensuremath{_{\mathnormal{#1}}}}}

\newcommand{\mthsup}[1]{\argemp{#1}{\ensuremath{^{\mathnormal{#1}}}}}

\newcommandx{\mth}[5][1=, 2=0, 4=, 5=]{{\ensuremath{\mthfnt[#1][#2]{#3}\mthsub{#4}\mthsup{#5}}}}

\newcommandx{\mtharg}[6][1=, 2=0, 4=, 5=]{{\mth[#1][#2]{#3}[#4][#5]\ensuremath{\argint{(}{#6}{)}}}}

\newcommand{\mthstyfun}{0}
\newcommand{\mthfun}[1][]{\mth[Fun][\argdef{#1}{\mthstyfun}]}

\definecolor{almond}{rgb}{0.94, 0.87, 0.8}
\definecolor{beige}{rgb}{0.96, 0.96, 0.86}

\newcommand{\ignore}[1]{}

\newcommand{\wrt}{wrt\ }

\makeatletter
\newcommand{\nobrackettag}[0]{\def\tagform@##1{\maketag@@@{##1}}}
\makeatother

\newcommand{\ltl}{\ensuremath{\textsc{ltl}}\xspace}

\newcommand{\ltlbcq}{\tcq\xspace}

\newcommand{\prop}{\ensuremath{\textsc{prop}}}
\newcommand{\props}{\ensuremath{\mthfun{props}}}
\newcommand{\ltlop}[1]{\operatorname{\mathbf{#1}}}
\newcommand{\ltlX}{\ltlop{X}}

\newcommand{\ltlU}{\ltlop{U}}

\newcommand{\tnot}{\neg}

\newcommand{\tup}[1]{\ensuremath{(#1)}\xspace}
\newcommand{\card}[1]{\ensuremath{|#1|}\xspace}

\newcommand{\A}{\ensuremath{\mathcal{A}}\xspace}
\newcommand{\I}{\ensuremath{\mathcal{I}}\xspace}
\newcommand{\J}{\ensuremath{\mathcal{J}}\xspace}

\renewcommand{\L}{\ensuremath{\mathcal{L}}\xspace}

\newcommand{\set}[1]{\{#1\}}
\newcommand{\bigset}[1]{\big\{#1\big\}}

\newcommand{\seqI}{\mathfrak{I}}
\newcommand{\seqP}{w\xspace}

\newcommand{\ra}{\ensuremath{\rightarrow}\xspace}

\newcommand{\cost}{\ensuremath{\mthfun{c}}\xspace}
\newcommand{\apply}{\ensuremath{\mthfun{mod}}\xspace}
\newcommand{\emptymods}{\ensuremath{\varepsilon}\xspace}
\newcommand{\emptytmods}{\ensuremath{\varepsilon}\xspace}

\newcommand{\ins}[1]{\operatorname{ins}(#1)\xspace}
\newcommand{\rem}[1]{\operatorname{rem}(#1)\xspace}

\newcommand{\delsym}{\operatorname{del}\xspace}
\newcommand{\del}{\delsym\xspace}
\newcommand{\addsym}{\operatorname{add}\xspace}
\newcommand{\add}[1]{\ensuremath{(\addsym,#1)}\xspace}
\newcommand{\fixsym}{\operatorname{fix}\xspace}
\newcommand{\fix}[1]{\ensuremath{(\fixsym,#1)}\xspace}

\newcommand{\op}{m\xspace}
\newcommand{\sep}{\operatorname{\cdot}\xspace}
\newcommand{\emptyA}{\A_\emptyset}

\newcommand{\mods}{\ensuremath{\mu}}
\newcommand{\tmods}{\ensuremath{\eta}}

\newcommand{\trans}[1]{\ensuremath{\stackrel{#1}{\ra}}}
\newcommand{\word}{\ensuremath{\Psi}\xspace}

\newcommand{\rset}{\ensuremath{\mathsf{N_R}}\xspace} 
\newcommand{\cset}{\ensuremath{\mathsf{N_C}}\xspace} 
\newcommand{\iset}{\ensuremath{\mathsf{N_I}}\xspace} 

\newcommand{\sig}{\mathsf{sig}}
\newcommand{\defequal}{:=}

\newcommand{\alc}{\ensuremath{\mathcal{ALC}}\xspace} 
 
\newcommand{\Amc}{\ensuremath{\mathcal{A}}\xspace}  
\newcommand{\Imc}{\ensuremath{\mathcal{I}}\xspace}   
\newcommand{\Jmc}{\ensuremath{\mathcal{J}}\xspace}   
\newcommand{\Kmc}{\ensuremath{\mathcal{K}}\xspace} 
\newcommand{\Tmc}{\ensuremath{\mathcal{T}}\xspace}
\newcommand{\assert}{\ensuremath{\alpha}\xspace}

\newcommand{\TKB}{TKB\xspace}
\newcommand{\tkb}{\ensuremath{\Gamma}\xspace}
\newcommand{\Aseq}{\ensuremath{\Lambda}\xspace}
\newcommand{\emptyAseq}{\ensuremath{\Aseq_\epsilon}\xspace}

\newcommand{\ind}{\mathsf{Ind}}

\newcommand{\query}{\ensuremath{\phi}\xspace}
\newcommand{\bquery}{\ensuremath{\phi}\xspace}
\newcommand{\pquery}{\ensuremath{\hat{\query}}\xspace}

\newcommand{\tquery}{\ensuremath{\varphi}\xspace}
\newcommand{\ptquery}{\ensuremath{\hat{\tquery}}\xspace}
\newcommand{\tformula}{\ensuremath{\varphi}\xspace}

\newcommand{\vset}{\ensuremath{\mathsf{N_V}}\xspace} 
\newcommand{\varsq}{\ensuremath{\mathsf{Var}}\xspace}
\newcommand{\fvarsq}{\ensuremath{\mathsf{FVar}}\xspace}
\newcommand{\indsq}{\ensuremath{\mathsf{Ind}}\xspace}
\newcommand{\match}{\ensuremath{h}\xspace}
\newcommand{\cans}[2]{\ensuremath{cert_{#1}(#2)}\xspace}

\newcommand{\hphism}{h}
\newcommand{\disj}{d}
\newcommand{\bis}{\beta}

\newcommand{\cq}{CQ}
\newcommand{\CQ}{\ensuremath{\textsc{cq}}\xspace}
\newcommand{\bcq}{BCQ}
\newcommand{\BCQ}{\ensuremath{\textsc{bcq}}\xspace}
\newcommand{\PBCQ}{\ensuremath{\textsc{pbcq}}\xspace}

\newcommand{\bool}{\ensuremath{\mathcal{B}}\xspace}

\newcommand{\bcBCQ}{\ensuremath{\bool(\BCQ)}\xspace}
\newcommand{\bcCQ}{\ensuremath{\bool(\CQ)}\xspace}
\newcommand{\tcq}{\ensuremath{\textsc{tcq}}\xspace}

\newcommand{\buchi}{B\"{u}chi\xspace}
\newcommand{\PA}{\ensuremath{P}\xspace}
\newcommand{\col}{\ensuremath{\mthfun{col}}\xspace}
\newcommand{\Acc}{\ensuremath{\mthfun{Acc}}\xspace}
\newcommand{\Col}{\ensuremath{Col}\xspace}
\newcommand{\painf}{\ensuremath{\mthfun{inf}}\xspace}

\newcommand{\ppath}{\rho\xspace}
\newcommand{\DFA}{\ensuremath{D}\xspace}
\newcommand{\MMG}{\ensuremath{G}\xspace}
\newcommand{\atmods}{\ensuremath{\mthfun{Atm}}\xspace}

\newcommand{\chformula}{\ensuremath{\chi}\xspace}
\newcommand{\alphabet}{\ensuremath{Al}\xspace}

\newcommand{\argmin}{\mathop{\mathrm{argmin}\xspace}}
\newcommand{\weight}{\mathop{\mthfun{w}\xspace}}

\begin{document}

\begin{frontmatter}

\title{Optimal Alignment of Temporal Knowledge Bases}

\author[A,C]{\fnms{Oliver}~\snm{Fern\'andez Gil}
\thanks{Corresponding Author. Email: oliver.fernandez@tu-dresden.de}
}
\author[B]{\fnms{Fabio}~\snm{Patrizi}
\thanks{Email: patrizi@diag.uniroma1.it}}
\author[B]{\fnms{Giuseppe}~\snm{Perelli}
\thanks{Email: perelli@di.uniroma1.it}} 
\author[A,C]{\fnms{Anni-Yasmin}~\snm{Turhan}
\thanks{Email: anni-yasmin.turhan@tu-dresden.de}} 

\address[A]{Theoretical Computer Science, TU Dresden, Germany}
\address[B]{Sapienza University of Rome}
\address[C]{Center for Scalable Data Analytics and Artificial Intelligence (ScaDS.AI) Dresden/Leipzig, Germany}

\begin{abstract}
%
%
%
%
Answering temporal CQs over temporalized Description Logic knowledge bases (TKB) is a main technique to realize ontology-based situation recognition. In case the collected data in such a knowledge base is inaccurate, important query answers can be missed. 
In this paper we introduce the TKB Alignment problem, which computes a variant of the TKB that minimally changes the TKB, but entails the given temporal CQ and is in that sense (cost-)optimal. We investigate this problem for \alc TKBs and conjunctive queries with LTL operators and devise a solution technique to compute (cost-optimal) alignments of TKBs that extends techniques for the alignment problem for propositional LTL over finite traces.
\end{abstract}

\end{frontmatter}


\section{Introduction}

Observing complex systems over time and drawing conclusions about their behavior 
is a core task for many AI systems. In particular, adaptive systems have to recognize situations in which an adaptation is useful. A well-investigated approach to do this is ontology-based situation recognition \cite{BBKTT-KIJ-20,AKRWZ-AIJ-21,Optique-17}. This approach is usually realized by modeling the observed system by a temporal knowledge base (TKB), where the data from the observed system is collected over time and stored in a sequence of ABoxes and a TBox that models important notions from the application domain. 
The situation to be recognized by the system is then modeled by a temporal query that will be answered over the sequence of ABoxes and the TBox. The situation recognition is then to detect predefined situations that are formalized as temporal (conjunctive) queries over the observed and enriched ABox sequence.
As in classical ontology-mediated query answering \cite{BO-RW-15}, the TBox enriches the data in the ABox sequence as it restricts its interpretation and allows for more conclusions. 
The semantics of  TKBs is given by an infinite  sequence of first-order interpretations. TKBs can be queried by temporal conjunctive queries (TCQs), which combine \ltl with conjunctive queries. Methods for answering temporal queries over TKBs and testing entailment of Boolean TCQs have been intensively investigated (\cite{BaaderBL15,BoLT-JWS15,AKRWZ-AIJ-21}).

Now, in many applications, the data is  collected from several sources and need not always be accurate. Consider the medical domain, where deviations of classical symptoms are frequent for certain patient groups or where examination methods such as blood test results can be inaccurate or discretized unsuitably. Thus the query need not return the expected answer, although the patient or, in the general case, the observed system is in a critical state that requires adaptation. The problem is to find a version of the TKB that admits to detect \enquote{near misses}. 

There are mainly two approaches developed to address the problem of errors or inaccuracies in DL knowledge bases. In case of inconsistent TKBs, ontology repairs restore consistent versions by deleting statements from the ABoxes \cite{BouKooTur2019,Bien-KIJ-20}. In case that information is missing in the ABoxes for the query to return answers, ABox abduction, i.e., adding new statements to the ABoxes has been investigated---mostly in the atemporal setting \cite{DR-AAAI-19,KoDeToSc-KR-20}.

In this paper, we investigate the new task of TKB Alignment, i.e., to modify the sequence of ABoxes by deletions or additions of statements so as to yield answers for the TCQ. 
Surprisingly, this problem has not been addressed in the literature yet. The goal of this paper is to develop an approach to solve instances of this new problem.

The well-known problem of Trace Alignment realizes a very similar task to TKB alignment: for a finite trace of observations  and a property specification expressed in Linear Temporal Logic (\ltl), a minimal modification of the 
trace is produced that satisfies the specification. This task has been extensively studied by the 
Business Process (BP) and AI communities, leading to effective solutions and implemented tools; 
see, e.g.,~\cite{DMMP17,Leoni2012,LeoniMA15}.
In all these settings, the observations recorded in a trace are propositional, i.e., each time point of the trace 
represents one of finitely many possible observables, modeled as propositions.


In this paper, we address the problem of 
TKB Alignment as a Trace Alignment problem in a much richer setting, where 
observables 
are described by DL concepts and roles, and properties are specified by a  temporalized query using DL atoms. Furthermore, the open world semantics of DLs is adopted, since entailment is considered instead of satisfaction as in classical propositional trace alignment.

We investigate the following setting for TKB alignment: a TCQ 
using (future) \ltl operators and a TKB written in the DL \alc, 
together with a  cost measure for edit operations on the ABox sequence. 
Solving TKB alignment is then to compute an ABox sequence  which, together with the TBox, entails the Boolean TCQ,
while guaranteeing cost-optimality of the modification. Intuitively, the cost-optimal version of the TKB states which minimal changes of the TKB would result in answers to the TCQ.

The technique we develop builds on an approach for deciding temporal query entailment over TKBs by \cite{BaaderBL15} 
and one for solving \ltl Trace Alignment for finite traces by \cite{DMMP17}, and extends them non-trivially.  
Our technique extends the former approach from verification to synthesis 
and the latter from the propositional to the DL setting, from propositional traces to TKBs, and from finite to infinite traces.
%
This ultimately results in an effective solution approach which can assess the deviation of irregular observations wrt standard ones
and define corrective actions to recover a standard observation.

%
%

\smallskip \noindent The detailed proofs for all results 
are supplied in the Appendix.

%

\section{Preliminaries}

In this section we recap basic notions on description logics, LTL, and trace alignment.

\subsection{Description Logic Knowledge Bases}

Description Logics (DLs) are a family of formal languages for representing 
knowledge and reasoning about it. In this work, we focus on the DL 
\alc~(\cite{Schmidt-SchaussS91}).

We fix three countably infinite sets of names: $\cset$ for concepts, $\rset$ for roles and $\iset$ for individuals.
\emph{Concepts} in \alc are defined inductively as follows:
\[C \defequal A \mid \neg C \mid C \sqcup C \mid \exists r.C \mid \top,\]
where $A \in \cset$, $r \in \rset$, 
$\top$ is the \emph{top-concept}  and  $\bot$ the \emph{bottom-concept}. 
We use the following standard abbreviations: $C \sqcap D$ for $\neg (\neg C \sqcup \neg D)$, $\forall r.C$ for $\neg (\exists r.\neg C)$, and $\bot$ for $\neg \top$.


DL concepts are interpreted over (first-order, FO) interpretations. 
An \emph{interpretation} $\Imc = (\Delta^\Imc, \cdot^\Imc)$ consists of a
domain $\Delta^\Imc$ and a function $\cdot^\Imc$ mapping each concept name 
$A$ to a set $A^\Imc\subseteq\Delta^\Imc$, each role name $r$ 
to a binary relation $r^\Imc\subseteq\Delta^\Imc \times \Delta^\Imc$, 
and each individual name $a$ to an element $a^\Imc\in\Delta^\Imc$. 
Based on this, the semantics of (complex) concepts is defined as follows:
$(\neg C)^\Imc =  \Delta^\Imc \setminus C^\Imc$, 
$(\exists r.C)^\Imc = \{d  \in \Delta^\Imc \mid \exists e. ((d,e) \in r^\Imc \wedge e \in C^\Imc)\}$, 
$(C \sqcup D)^\Imc =  C^\Imc \cup D^\Imc$, and
$\top^\Imc = \Delta^\Imc$.

\emph{General concept inclusions} (GCIs) are statements 
of the form $C \sqsubseteq D$, expressing 
inclusion relationships between concepts. 
A \emph{TBox} (denoted \Tmc) is a finite set of GCIs. 
A \emph{model} of a TBox \Tmc is an interpretation \Imc that satisfies all 
GCIs in \Tmc, i.e., for all $C \sqsubseteq D \in \Tmc$, it holds that 
$C^\Imc \subseteq D^\Imc$.
A TBox is \emph{satisfiable} if it has a model.

Statements $A(a)$ and $r(a,b)$ are called, respectively,
\emph{concept assertion} and 
\emph{role assertion},
where $a, b \in \iset$, 
$A \in \cset$ and 
$r\in\rset$.
An interpretation $\Imc$ satisfies $A(a)$
if $ a^\Imc \in A^\Imc$, and 
satisfies $r(a,b)$
if $(a^\Imc, b^\Imc)  \in r^\Imc$.
An \emph{ABox} \Amc is a finite set of (concept or role) assertions. 
An interpretation \Imc is a \emph{model} of an ABox \Amc, 
if \Imc satisfies all assertions in \Amc.

A \emph{DL knowledge base} (KB) is a pair $\Kmc = (\Tmc, \Amc)$, with $\Tmc$ a TBox and 
$\Amc$ an ABox. 
An interpretation \Imc is a \emph{model} of $\Kmc = (\Tmc, \Amc)$, written 
$\Imc\models\Kmc$, if \Imc is a model of \Tmc and \Amc. 
A KB is \emph{consistent} if it has a model.

\subsection{Conjunctive Queries}\label{sub:sec:cqs}
Prominent reasoning problems investigated in the last decade concern 
\emph{conjunctive queries}.
We briefly recap related definitions and results.
\begin{definition}[Conjunctive query]\label{def:conjunctive:queries}
   	    Let $\vset$ be a set of variables. 
   	    A \emph{conjunctive query (\cq)} is an expression of the form 
   	    $\query=\exists \bar{y}.\psi$, where $\bar{y}$ is a  
   	    tuple of variables from $\vset$ and $\psi$ is a finite conjunction of 
   	    \emph{atoms} of the form:
   	    	$A(z)$, for $A \in \cset$ and $z \in \vset \cup \iset$, or 
   	    	$r(z,z')$, for $r\in \rset$ and $z,z' \in \vset \cup \iset$.
\end{definition}
By $\CQ$ we denote the set of all \cq s (over $\cset$, $\rset$, $\iset$, 
$\vset$). We write $A(z)\in \query$ to state that atom $A(z)$ occurs in 
$\query$, and likewise for $r(z,z')$.

In this work, we combine \cq s using Boolean connectives. 
\begin{definition}[Boolean combination of \cq s]
	A formula $\bquery$ is a \emph{Boolean combination of \cq s} iff:
	\[\phi \defequal \bquery'\mid \neg \bquery\mid \bquery \lor \bquery, 
		\text{ where }\bquery'\in\CQ.\]
       %
       %
   \end{definition}
   As standard, $\bquery_1\land\bquery_2$ abbreviates
   $\lnot(\neg \bquery_1\lor \neg\bquery_2)$.
Given a Boolean combination of \cq s $\bquery$, we denote by $\varsq(\bquery)$, $\fvarsq(\bquery)$ and $\indsq(\bquery)$ the set of variables, free variables and individual names occurring in $\bquery$, respectively. 
A query with no free variables is called \emph{Boolean}, whereas a query with $\ind(\bquery)=\emptyset$ is called \emph{pure}.
$\BCQ$ denotes the set of \emph{Boolean \cq s} (\bcq s), 
$\bcCQ$ the set of Boolean combinations of \cq s, 
and $\bcBCQ$ the set of Boolean combinations of \bcq s.

The semantics of \bcq s is defined in terms of a 
satisfaction relation between interpretations and \bcq s.
 \begin{definition}[Semantics of \bcq s]\label{def:sem:cqs}
 	An interpretation $\Imc$ is \emph{a model} of (or \emph{satisfies}) 
 	a {\bcq} $\query$, written $\Imc \models \query$, iff
 	there exists a mapping $\match:\varsq(\query) \cup \indsq(\query) \rightarrow \Delta^\Imc$,
 	called a \emph{match}, s.t.:
 	\begin{itemize}
 		\item
 		$\match(a) = a^\Imc$ for all $a \in \indsq(\query)$;
 		\item
 		$\match(z) \in A^\Imc$ for all $A(z) \in \query$; and
 		\item
 		$(\match(z),\match(z')) \in r^\Imc$ for all $r(z,z') \in \query$.
 	\end{itemize}
    %
	\end{definition}
	These notions straightforwardly extend to $\bcBCQ$.
    \begin{definition}[Semantics of Boolean combinations of \bcq s]\label{def:sem:boolean:cqs}
 	An interpretation $\Imc$ is \emph{a model} of (or \emph{satisfies}) 
 	a query $\query \in \bcBCQ$, written $\Imc \models \query$, iff:
    \begin{itemize}
    	\item $\bquery\in\BCQ$ and $\Imc\models\bquery$; or
    	\item 
    	$\bquery=\lnot\bquery_1$ and $\Imc\not\models\bquery_1$; or    	
    	\item
    	$\bquery=\bquery_1 \lor \bquery_2$ and $\Imc \models \bquery_1$ or $\Imc \models \bquery_2$.
    \end{itemize}
 \end{definition}

A query $\bquery \in \bcBCQ$ is \emph{satisfiable} wrt a KB $\Kmc$, if $\Imc \models \bquery$ for some model $\Imc$ of $\Kmc$. A knowledge base \Kmc \emph{entails} a query $\bquery$
 (written $\Kmc \models \bquery$),  if $\Imc \models \bquery$ for all models $\Imc$ of $\Kmc$.
 
In case of non-Boolean queries, one is interested in computing the certain answers. More precisely, given a KB $\Kmc=(\Tmc,\Amc)$ and $\bquery \in \bcCQ$ with free variables $\bar{x}=(x_1,\ldots,x_k)$, a tuple $\bar{a}=(a_1,\ldots,a_k)$ of individuals in $\iset$ is a \emph{certain answer of $\bquery$ wrt\ $\Kmc$} if $\Kmc \models \bquery[\bar{a}]$, where $\bquery[\bar{a}]$ is the Boolean query obtained from $\bquery$ by replacing each occurrence of $x_i$ by $a_i$ ($1 \leq i \leq k$).
 %
 We denote by $\cans{\Kmc}{\bquery}$ the set of certain answers of $\bquery$ wrt\ $\Kmc$. If $\bquery$ is a Boolean query and $\Kmc \models \bquery$, then $\cans{\Kmc}{\bquery}=\{()\}$.
 
 The entailment problem for  \bcq s wrt\ $\alc$ knowledge bases is ExpTime-complete (\cite{Lutz08,OrtizSE08}). 
 It was shown in \cite{BaaderBL15} for $\bcBCQ$ that satisfiability of a conjunction of \cq-literals (i.e.\ either a Boolean \cq\ or a negated Boolean \cq) wrt\ $\alc$ knowledge bases is an ExpTime-complete problem.
 An easy consequence of this (and of $\bcBCQ$ being \emph{closed under negation}) is that satisfiability and entailment of arbitrary Boolean combinations of \bcq s w.r.t.\ $\alc$ knowledge bases are also ExpTime-complete problems.
 

\subsection{Propositional Linear Temporal Logic}
The kind of properties we focus on in this paper concerns the evolution of 
a knowledge base over time. To express relevant properties, we need a temporal logic. 
We review the basics on propositional Linear Temporal Logic (\ltl), which 
will be later lifted to \cq s and used on to address TKB Alignment.

The language of Linear Temporal Logic (\ltl) formulas $\tquery$ is defined over a finite set of propositions $\prop$, as follows: 
\[
\tquery \defequal p \mid \tnot \tquery \mid \tquery \vee \tquery \mid \ltlX \tquery \mid \tformula \ltlU \tquery
		\text{,~ with $p \in \prop$.}
\]
The set $\props(\tquery)$ denotes the finite set of propositions occurring in $\tquery$.
$\ltl$ formulas are interpreted over infinite words, also called~\emph{(propositional) traces}, 
$\seqP = \seqP_0 \seqP_1 \cdots \in{(2^\prop)}^\omega$.
\begin{definition}[$\ltl$ semantics]\label{def:ltl-semantics}
	Given a formula $\tquery\in\ltl$, a trace $\seqP=\seqP_0 \seqP_1\cdots$, and an index $i$, 
	we inductively define when $\seqP, i$ \emph{satisfy} $\tquery$, denoted $\seqP, i \models \tquery$, as follows:
		\begin{itemize}
			\item 
			$\seqP, i \models p$, if $p\in \seqP_i$;	
			\item
			$\seqP, i \models \tnot \tquery$, if $\seqP, i \not\models \tquery$;
			\item
			$\seqP, i \models \tquery_{1} \vee \tquery_{2}$, if $\seqP, i \models \tquery_{1}$ or $\seqP, i \models \tquery_{2}$;
			\item
			$\seqP, i \models \ltlX \tquery$, if $\seqP, i + 1 \models \tquery$;
			\item
			$\seqP, i \models \tquery_{1} \ltlU \tquery_{2}$ if there exists $j \geq i$ s.t.~$\seqP, j \models \tquery_{2}$ and $\seqP, k \models \tquery_{1}$, 
			for $k=i,\ldots,j - 1$.
		\end{itemize}
	We say that $\seqP$ \emph{satisfies} $\tquery\in\ltl$, written $\seqP \models \tquery$, iff $\seqP,0\models\tquery$.
\end{definition}
We denote the set of traces satisfying $\tquery$ as $\L(\tquery)=\set{\seqP\in{(2^\prop)}^\omega\mid\seqP\models\tquery}$.
It is well-known that for every $\tquery\in\ltl$ there exists a
\emph{deterministic parity automaton} (DPA) $\PA_\tquery$ accepting exactly 
$\L(\tquery)$.

A \emph{deterministic parity automaton}
is a tuple $\PA=(\alphabet,Q,\delta,q_0,\col)$, where:
$\alphabet$ is the finite input alphabet, $Q$ is the finite set of states, 
$\delta: Q\times\alphabet\ra Q$ is the transition function, 
$q_0\in Q$ is the initial state, and
$\col:Q\ra\Col$ is a 
\emph{coloring function}, mapping the states of \PA into a finite set of \emph{colors}
$\Col\subset\mathbb{N}_0$.
DPAs are similar to deterministic finite-state automata (DFA),
but accept \emph{infinite} traces and thus have a different accepting condition.

For a DPA $\PA$, a \emph{finite run from state $q\in Q$} is a sequence
$\ppath=q\trans{\seqP_0}q_1\trans{\seqP_1}\cdots\trans{\seqP_{n-1}}q_n$ s.t.~$\delta(q,\seqP_0)=q_{1}$ and 
$\delta(q_i,\seqP_i)=q_{i+1}$, for $0<i<n$. 
We define \emph{infinite} runs analogously, 
for $n=\infty$. Unless stated otherwise, runs are always infinite and start in the initial state  $q_0$ of $\PA$.

Given a run $\ppath$ of $\PA$, let $\painf_Q(\ppath,\PA)$ be the set of states occurring infinitely 
many times in $\ppath$. Obviously, $\painf_Q(\ppath,\PA)\neq\emptyset$ iff $\ppath$ is infinite.
Let $\painf(\ppath,\PA)=\set{\col(q)\in\Col \mid q\in \painf_Q(\ppath,\PA)}$ be  
the set of colors ``visited'' infinitely many times by $\ppath$.
A run $\ppath$ from a state $q\in Q$ is \emph{accepting} iff
$\min\set{\painf(\ppath,\PA)}$ is even.
When this is the case, $q\in Q$ is an \emph{accepting} state. 
By $\Acc(\PA)$, we denote the set of all accepting states of $\PA$ and call $\Acc(\PA)$ the \emph{accepting}
set of $\PA$.

\begin{lemma}[\cite{KingKV01}]
The accepting set $\Acc(\PA)$ of a DPA $\PA=(\alphabet,Q,\delta,q_0,\col)$ can be computed in time
$(\card{Q}+\card{\delta})\log\card{\Col}$, where
$\card{\delta}=\card{\set{(q,q')\in Q\times Q\text{ s.t.~}q'=\delta(a,q)\text{, for some }a}}$.
\end{lemma}

For a DPA $\PA=(\alphabet,Q,\delta,q_0,\col)$ and a trace 
$\seqP=\seqP_0\seqP_1\cdots$, the (unique) run \emph{induced} by $\seqP$ is the 
run $\ppath=q_0\trans{\seqP_0}q_1\trans{\seqP_1}\cdots$. 
A trace $\seqP$ is \emph{accepted} by $\PA$ iff the run $\ppath$ induced by 
$\seqP$ is accepting.
By $\L(\PA)$ we denote the \emph{language} of $\PA$, i.e., the set of all traces  accepted by $\PA$.

\begin{theorem}[\cite{Var95,Pit07}]\label{thm:sat-bstates}
For every $\tquery\in\ltl$ there exists a DPA $\PA_\tquery=(2^{\prop},Q,\delta,q_0,\col)$ 
s.t.~$\L(\PA_\tquery)=\L(\tquery)$.
$\PA_\tquery$ can be computed in doubly exponential time and has doubly exponential size wrt $\tquery$.
\end{theorem}



\subsection{Entailment of Temporal Conjunctive Queries}
TKB Alignment is closely related to 
checking \emph{temporal query entailment} (TQE), studied in~\cite{BaaderBL15}.
We  briefly recall the main definitions and results from that work, 
possibly adapted to our setting.
\begin{definition}[Temporal Knowledge Base (TKB)]\label{def:tkb}
	A \emph{temporal knowledge base} (TKB)  is a pair 
	$\tkb = (\Tmc, \Aseq)$, where $\Tmc$ is a TBox and $\Aseq= \Amc_0 \cdots \Amc_\ell$ is
	a finite sequence of ABoxes.
\end{definition}
A \emph{FO trace} is an infinite sequence $\seqI=\Imc_0 \Imc_1 \cdots$ 
of interpretations $\Imc_i=(\Delta,.^{\Imc_i})$ over a fixed domain $\Delta$.
\begin{definition}[Model of a TKB]\label{def:tkb-model}	
    Given a TKB $\tkb = (\Tmc, \Amc_0\cdots\Amc_\ell)$,
    a FO trace $\seqI=\Imc_0\Imc_1\cdots$ is a \emph{model} of $\tkb$, iff:
	$\I_i \models \Amc_i$, for $0\leq i \leq\ell$; and
	$\I_i \models \Tmc$ for all $i \geq 0$.
	If these conditions hold, $\seqI$ \emph{satisfies} $\tkb$, written $\seqI \models \tkb$.
\end{definition}
Next, we introduce the language \ltlbcq of \emph{(simple) temporal conjunctive queries} 
(TCQ), which essentially lifts propositions in \ltl to \bcq s.
$\ltlbcq$ formulas are obtained as in $\ltl$, by replacing $\prop$ 
with a finite set $\PBCQ\subset\BCQ$:
	\[
		\tquery \defequal \query \mid \tnot \tquery \mid \tquery \vee \tquery \mid \ltlX \tquery \mid \tquery \ltlU \tquery
		\text{, ~with $\query \in \PBCQ$.}
    \]

$\ltlbcq$ formulas  are evaluated over FO traces.

\begin{definition}[\tcq Semantics]~\label{def:tcq-semantics}
	Given a formula $\tquery \in \ltlbcq$, an infinite trace $\seqI=\I_0 \I_1 \cdots$, and an index $i$, 	
	we inductively define when $\seqI$ \emph{satisfies} $\tquery$ from $i$, denoted $\seqI, i \models \tquery$, as follows:
	\begin{itemize}
		\item 
		$\seqI, i \models \query$, if $\Imc_i \models \query$, for $\query \in \PBCQ$;
		\item
		$\seqI, i \models \tnot \tquery$ if $\seqI, i \not\models \tquery$;
		\item
		$\seqI, i \models \tquery_{1} \vee \tquery_{2}$, if $\seqI, i \models \tquery_{1}$ or $\seqI, i \models \tquery_{2}$;
		\item
		$\seqI, i \models \ltlX \tquery$, if $\seqI, i + 1 \models \tquery$;
		\item
		$\seqI, i \models \tquery_{1} \ltlU \tquery_{2}$ if there exists $j \geq i$ s.t.~$\seqI, j \models \tquery_{2}$ and $\seqI, k \models \tquery_{1}$, 
		for $k=i,\ldots,j - 1$.
	\end{itemize}
	We say that $\seqI$ \emph{satisfies} $\tquery$, written $\seqI \models \tquery$, if $\seqI,0\models\tquery$.
\end{definition}
The semantics of temporal operators for TCQs is analogous to that of $\ltl$ (see Def.~\ref{def:ltl-semantics}); however, for TCQs the base case accounts for the satisfaction of \bcq s by the FO interpretations occurring in the trace.

We observe that the variant of TCQs we use here differs from that introduced in~\cite{BaaderBL15}, in that we disallow \emph{past} operators~(\cite{Gabbay87}). 
However, such restriction comes without loss of generality.
This is because the two semantic variations have essentially the same expressive power, as future operators have the ability to mimic the past ones~(\cite{Wilke99,KupfermanPV01}).

The notion of \emph{temporal conjunctive query entailment} used here is the 
same as that in~\cite{BaaderBL15}, once past operators are disallowed in TCQs.
\begin{definition}[TCQ Entailment]\label{def:tcq-entailment}
	Given a TKB $\tkb$ and a TCQ $\tquery \in \ltlbcq$. The TKB $\tkb$ 
	\emph{entails} $\tquery$, written $\tkb \models \tquery$, 
	iff $\seqI \models \tquery$,
	for every model $\seqI$ of \tkb.
\end{definition}
Checking TCQ entailment (TQE) 
is the problem of deciding whether $\tkb \models \tquery$.

\section{The TKB Alignment Problem}
\label{sec:tkb-align}

We generalize the verification problem of TQE
to a \emph{synthesis} version, consisting
in \emph{minimally} modifying the sequence 
$\Aseq$ of a TKB $\tkb=(\Tmc,\Aseq)$, 
to obtain a TKB $\tkb'=(\Tmc,\Aseq')$, s.t.~$\tkb'\models\tquery$. 
Observe that if $\tkb\models\varphi$, the problem amounts to checking TQE.
To define the problem, we  formalize next the notions of ABox- and TKB-modification, and minimality.

To modify the ABoxes occurring in a TKB, we consider two kinds of ABox operations, namely 
\emph{insertion} and \emph{removal} of a (concept or role) assertion $\assert$, 
respectively denoted as $\ins{\alpha}$ and $\rem{\alpha}$.
The result of applying such operations to an ABox $\Amc$ is given by the function $\apply$:
\begin{inparaenum}[(\it i)]
    \item $\apply(\ins{\assert}, \Amc) =\Amc\cup\set{\assert}$;
    \item $\apply(\rem{\assert}, \Amc) =\Amc\setminus\set{\assert}$.
\end{inparaenum}
An \emph{ABox-modification} is a (possibly empty) sequence $\mods=\op_0\cdots \op_n$ of ABox operations 
$\op_i$. By $\emptymods$ we denote the empty ABox-modification.
The semantics of applying an ABox-modification $\mods$ to an ABox $\Amc$ is obtained by 
inductively extending $\apply$ to sequences of operations: 
\begin{inparaenum}[(\it i)]
	\item $\apply(\emptymods, \Amc) = \Amc;$
	\item $\apply(\op\sep\mods,\Amc) = \apply(\mods,\apply(\op,\Amc))$, where $\sep$ is the concatenation operator.
\end{inparaenum}

We assume every operation $\op$ has a strictly positive cost $\cost(\op)\in\mathbb{R^+}$.
The \emph{cost} of an ABox-modification 
$\mods=\op_1\cdots \op_n$ is defined as 
$\cost(\mods) :=\sum_{i=1}^n \cost(\op_i)$, with $\cost(\emptymods)=0$.

In addition to modifying its ABoxes, 
a TKB can be modified by adding or removing ABoxes.
Let $\atmods=\set{\fix{\mods},\add{\mods}, \del \mid \mods \text{ is an ABox-modification}}$
be the set of \emph{atomic TKB-modifications}. 
Intuitively, 
$\fix{\mods}$ stands for the modification of an ABox through the application of $\mods$,
$\add{\mods}$ for the addition of the empty ABox followed by the application of $\mods$, 
and $\del$ for ABox deletion. 

A \emph{TKB-modification} is a finite sequence $\tmods$ of atomic TKB-modifications, with $\emptymods$ denoting the empty TKB-modification.
Notice that,  by a slight abuse of notation, we use $\emptymods$ to denote both the empty ABox-modification and the empty TKB-modification; the intended 
meaning is clear from the context.

The result of applying a TKB-modification 
$\tmods$ to a sequence 
$\Aseq=\Amc_0\cdots\Amc_\ell$ is the sequence $\apply(\tmods,\Aseq)$, 
inductively defined as follows, where 
$\emptyAseq$ denotes the empty sequence of ABoxes,
$\emptyA$ the empty ABox, and $\Amc|\Aseq=\Amc\Amc_0\cdots\Amc_\ell$:

\begin{itemize}
    \item $\apply(\varepsilon,\Aseq) = \Aseq$;
  
    \item $\apply(\fix{\mods}, \emptyAseq) = \emptyAseq$;
  
    \item $\apply(\fix{\mods} \sep \tmods, \Amc|\Aseq) = \apply(\mods,\Amc)|\apply(\tmods,\Aseq)$;
    \item $\apply(\add{\mods}\sep\tmods, \Aseq) = \apply(\mods,\emptyA)|\apply(\tmods,\Aseq)$;
	
	\item $\apply(\del \sep \tmods, \emptyAseq)) = \apply(\tmods, \emptyAseq)$;
	
	\item $\apply(\del \sep \tmods, \Amc|\Aseq) = \apply(\tmods, \Aseq)$.
\end{itemize}
For a given TKB $\tkb = (\Tmc, \Aseq)$ and a TKB-modification $\tmods$, 
we define $\apply(\tmods,\tkb) = (\Tmc, \apply(\tmods, \Aseq))$.
The cost function naturally extends to TKB-modifications and ABox sequences:
\begin{itemize}
	\item $\cost(\emptytmods, \Aseq) = 0$;
	\item $\cost(\fix{\mods} \sep \tmods, \emptyAseq) = \cost(\tmods,\emptyAseq)$;
	\item $\cost(\fix{\mods} \sep \tmods, \Amc|\Aseq) = \cost(\mods) + \cost(\tmods,\Aseq)$;
	\item $\cost(\add{\mods} \sep \tmods, \Aseq) = 1+ \cost(\mods) + \cost(\tmods,\Aseq)$;
	\item $\cost(\del \sep \tmods, \emptyAseq) = \cost(\tmods, \emptyAseq)$;
	\item $\cost(\del \sep \tmods, \Amc|\Aseq) = \big(\sum_{\assert \in \Amc} \cost(\rem{\assert})\big) + 1 + \cost(\tmods, \Aseq)$.
\end{itemize}
The cost of $\add{\mods}$ is that of adding the empty ABox (taken as 1) and applying $\mods$ to it; similarly, the cost of $\del$ is that of emptying 
the ABox, by removing all of its assertions, and removing the resulting empty ABox (i.e., 1).
For a TKB $\tkb=(\Tmc,\Aseq)$ and a TKB-modification $\tmods$, we let $\cost(\tmods,\tkb) =\cost(\tmods,\Aseq)$.

We can now introduce the TKB-alignment problem.
\begin{definition}[TKB Alignment] \label{def:tkb-alignment}
Given a \TKB $\tkb$ and a TCQ $\varphi$, 
the \emph{\TKB-alignment} problem
 consists in finding a minimal-cost TKB-modification $\tmods^*$ (if any) 
s.t.~$\apply(\tmods^*,\tkb) \models \varphi$.
\end{definition}

Observe that, for every TKB-modification $\tmods$ and TKB $\tkb$, 
it holds that $\apply(\tmods,\tkb) = \apply(\tmods \cdot \fix{\epsilon},\tkb)$ and
$\cost(\tmods, \tkb) = \cost(\tmods \cdot \fix{\epsilon}, \tkb)$.
That is, appending a sequence of $\fix{\epsilon}$ to $\tmods$ does not affect the result or
the cost of modifying $\tkb$.
Thus, we can always extend $\tmods$ 
to guarantee that the combined number of occurrences of 
deletions $\del$ and fixes $\fix{\mods}$ in $\tmods$ equals at least the number of ABoxes in $\Aseq$.
For technical convenience, from now on, we assume this is the case for every 
$\tmods$.

\subsection{Solving TKB Alignment}

Our solution approach consists in reducing TKB Alignment to Shortest Path. To this end, we construct a graph, called \emph{Minimal-instantiation Graph}, with each edge labelled by an atomic TKB-modification 
and its corresponding cost, s.t.~every shortest path from a suitably defined \emph{initial} node to one node from a (suitably defined) \emph{target} set, represents an optimal solution to the original TKB Alignment instance.
The construction of such a graph is based on several intermediate structures, which we present and discuss next.

Consider a TKB $\tkb=(\Tmc,\Aseq)$ and a TCQ $\tquery$. We start with the construction of a DPA intended to accept the set of models of $\tquery$.
To this end, we adopt an approach similar to that of~\cite{BaaderBL15}, which uses
the  \emph{propositional abstraction} of $\tquery$.
%
If we view every \bcq~$\query\in\PBCQ$ as a proposition $\pquery$, then $\tquery$ can be viewed as an \ltl formula.
This is called the \emph{propositional abstraction} of $\tquery$, denoted $\ptquery$.
Obviously, $\props(\ptquery)$ is the set of all propositions $\pquery$ occurring 
in $\ptquery$, each corresponding to exactly one BCQ $\query\in\PBCQ$.

Since $\ptquery\in\ltl$, we can now use the \buchi automaton (BA) construction 
of~\cite{Var95}, to obtain a BA that recognizes $\L(\ptquery)$ and then use the 
BA-to-DPA construction of~\cite{Pit07} to obtain the 
DPA $\PA_{\ptquery}=(2^{\props(\ptquery)},Q,q_0,\delta,\col)$ of $\ptquery$.
The importance of $\PA_{\ptquery}$ lies in the fact that, 
although it reads input words $\word=\Phi_0 \Phi_1 \cdots\in {(2^{\props(\ptquery)})}^\omega$
and not FO traces $\seqI=\Imc_0\Imc_1\cdots$, it fully characterizes $\L(\varphi)$, 
as discussed below.

For $\Phi\in 2^{\props(\ptquery)}$, let 
$\chformula(\Phi)=\bigwedge_{\pquery \in \Phi} \query \wedge \bigwedge_{\pquery \in \props(\ptquery) \setminus \Phi} \neg \query$.
Since the conjuncts in $\chformula(\Phi)$ are possibly negated \bcq s from $\PBCQ$, and
not propositional abstractions,
$\chformula(\Phi)\in\bcBCQ$.
If $\chformula(\Phi)$ is consistent, i.e., admits at least one model,
$\Phi$ is called a \emph{type}.
When $\Imc\models\chformula(\Phi)$, we call $\Phi$ the \emph{type} of $\Imc$. 
This notion naturally extends to traces by
defining the \emph{trace type} of a FO trace 
$\seqI=\Imc_0\Imc_1\cdots$ 
as the word $\word=\Phi_0\Phi_1\cdots\in{(2^{\props(\ptquery)})}^\omega$ 
s.t.~$\Phi_i$ is the type of $\Imc_i$, for all $i\geq 0$.
We have the following result.

\begin{restatable}{lemma}{typing}
	\label{lem:typing}
    Every FO interpretation has a unique type and every type admits an FO interpretation.
    Moreover, 
    every FO trace has a unique trace type and every trace type admits an FO trace.
\end{restatable}
The following result relates $\L(\PA_{\ptquery})$ and $\L(\tquery)$.
\begin{restatable}{theorem}{thmbridge}\label{thm:bridge}
Consider a TCQ $\tquery$.
For every FO trace $\seqI$ of type $\word$, 
it holds that $\seqI\in\L(\tquery)$ iff $\,\word\in\L(\PA_{\ptquery})$.
\end{restatable}

Since $\PA_{\ptquery}$ is independent of $\tkb$, it cannot 
be used to search for the desired minimal-cost modification.
For this, we can use a deterministic finite-state automaton (DFA) 
$\DFA$,
called the \emph{repair-template} DFA for $\tkb$ and $\tquery$.
The definition of $\DFA$ requires an auxiliary DPA, 
called the $\Tmc$-\emph{reduct} of $\PA_{\ptquery}$, 
to define the final states of $\DFA$.

\begin{definition}[$\Tmc$-reduct of $\PA_{\ptquery}$]\label{def:reduct}
Given a TBox $\Tmc$ and a TCQ $\tquery$, let 
$\PA_{\ptquery}=(2^{\props(\ptquery)},Q,\delta,q_0,\col)$. 
The $\Tmc$-reduct of $\PA_{\ptquery}$ 
is the DPA 
$\PA_{\ptquery}^{\Tmc}=(2^{\props(\ptquery)},Q^\Tmc,\delta^{\Tmc},q_0,\col^{\Tmc})$ 
s.t.:
\begin{itemize}
	\item $Q^\Tmc=Q\cup\set{q^*}$, with $q^*\notin Q$;
	\item $\delta^\Tmc(q,\Phi)=q'$, iff either:
		\begin{itemize}
			\item $\chformula(\Phi)$ is satisfiable wrt $\Tmc$ and $\delta(q,\Phi)=q'$; or
			\item $\chformula(\Phi)$ is not satisfiable wrt $\Tmc$ and $q'=q^*$; or
			\item $q=q'=q^*$;
		\end{itemize}
	\item $\col^\Tmc(q^*)=1$ and $\col^{\Tmc}(q)=\col(q)+1$, for all $q\in Q$.
\end{itemize}
\end{definition}

\begin{restatable}{lemma}{reductcomplexity}\label{lem:t-reduct-complexity}
The $\Tmc$-reduct of $\PA_{\ptquery}$ can be computed in doubly exponential time and has doubly exponential size wrt \tquery.
\end{restatable}

Intuitively, the $\Tmc$-reduct of $\PA_{\ptquery}$ accepts a 
trace type $\word$ iff there exists a FO trace $\seqI$ of type $\word$ 
that does not satisfy $\tquery$ and contains only interpretations satisfying $\Tmc$.
Let $\Acc(\PA^\Tmc_{\ptquery})$ be the acceptance set of $\PA^\Tmc_{\ptquery}$. We have the following.

\begin{restatable}{lemma}{lemtail}\label{lem:tail}
	
	Consider a finite sequence $\word'=\Phi_0 \cdots \Phi_{k - 1}$ of 
	types and 
	the finite run $\ppath=q_0 \trans{\Phi_0} \cdots \trans{\Phi_{k - 1}}q_k$ induced in $\PA_{\ptquery}$.
	Then,  
	$q_k \notin \Acc(\PA^\Tmc_{\ptquery})$ iff
	for every type $\word = \Phi_0 \cdots \Phi_{k - 1} \Phi_{k} 
	\Phi_{k + 1} \cdots$, having $\word'$ as a prefix, and for all 
	traces $\seqI = \Imc_0 \Imc_1 \cdots$ 
	of type $\word$, if, for all $i \geq k$, it holds that 
	$\Imc_i\models\Tmc$, then $\seqI \models \tquery$.
		
\end{restatable}
Recall that we are looking for a TKB $\tkb' = (\Tmc, \Aseq' = \Amc_0\cdots\Amc_{\ell'})$, 
obtained as modification of $\tkb=(\Tmc,\Aseq)$, 
s.t.~$\tkb' \models \tquery$, i.e., all models $\seqI$ of $\tkb'$ 
(Def.~\ref{def:tkb-model}) are s.t.~$\seqI \models \tquery$ (Def.~\ref{def:tcq-entailment}).
Lemma~\ref{lem:tail} implies that every model $\seqI$ of $\tkb'$ must belong to some trace type $\word_{\seqI}$ whose induced run in $\PA_{\ptquery}$ touches some $q_k \notin \Acc(\PA^\Tmc_{\ptquery})$, for $k = \ell' + 1$.
Based on this, we next define the \emph{repair-template} DFA.
\begin{definition}[Repair-template DFA]\label{def:rtdfa}
	Given a TKB $\tkb=(\Tmc,\Aseq)$ with $\Aseq=\Amc_0\cdots\Amc_\ell$ 
	and a TCQ $\tquery$, let $\PA_{\ptquery}^{\Tmc}=(2^{\props(\ptquery)},Q^\Tmc,\delta^{\Tmc},q_0,\col^{\Tmc})$ 
	be the $\Tmc$-reduct of $\PA_{\ptquery}=(2^{\props(\ptquery)},Q,\delta,q_0,\col)$.

	The \emph{repair-template DFA (RT-DFA) for $\tkb$ and $\tquery$} is the 
	DFA $\DFA=(\alphabet,S,s_0,\gamma,F)$ s.t.:
	\begin{itemize}
		\item $\alphabet= (\set{\fixsym,\addsym}\times2^{2^{\props(\ptquery)}})\cup\set{\delsym}$ is the alphabet;
		\item $S=2^Q\times\set{0,\ldots,\ell+1}$ is the set of states;
		\item $s_0=(\set{q_0},0)$ is the initial state;
		\item $\gamma: S \times \alphabet \ra S$ is the transition function 
		s.t.~$\gamma((Z,i),X)=(Z',i')$ iff either:
			\begin{enumerate}[leftmargin=*]
			    \item 
			        $X=\delsym$, $Z=Z'$, and $i'=\min\set{i + 1, \ell + 1}$; or
			\item all of the following hold:
				\begin{enumerate}
					\item\label{it:dfa-1} $X=(\sigma,\Upsilon)$, with $\sigma\in\set{\fixsym,\addsym}$;
					\item $q'\in Z'$ iff $\delta(q,\Phi)=q'$, for $q\in Z$ and $\Phi\in\Upsilon$;
					\item\label{it:dfa-2} there exists an ABox $\Amc$ consistent with $\Tmc$ s.t.:
					$(\Tmc,\Amc)\models 
	                        			\bigvee_{\Phi \in \Upsilon}\chformula(\Phi)\land
	                        			\bigwedge_{\Phi \not\in \Upsilon}\lnot\chformula(\Phi)$;
    			        		\item $i'=
							\begin{cases}
								\min\set{i+1,\ell+1}, \text{ if } \sigma=\fixsym\\
								i, \text{ if } \sigma=\addsym
							\end{cases}
						$
				\end{enumerate}
			\end{enumerate}
		\item $F = \set{(Z, \ell + 1) \in S \mid Z\cap\Acc(\PA_{\ptquery}^{\Tmc})=\emptyset}$ is
			the set of final states.
	\end{itemize}
\end{definition}

\begin{restatable}{lemma}{rtdacomplexity}\label{lem:rtda-complexity}
The RT-DFA for a TKB  $\tkb$ and a TCQ $\tquery$ can be computed in triply exponential time and has triply exponential size wrt $\tquery$.
\end{restatable}

Observe that the right-hand side expression of the entailment ($\models$) in
Item~\ref{it:dfa-2} above is a Boolean combination of \bcq s.
The purpose of the RT-DFA is to capture the solution space of TKB Alignment for 
$(\Tmc,\Aseq)$ and $\tquery$, in the following sense:
\begin{inparaenum}[(\it i)]
    \item from every accepted word $w$, some TKB modification $\tmods$ can be derived s.t.~$\apply(\tmods, \tkb) \models \tquery$,
    and
    \item every TKB modification $\tmods$ s.t.~$\apply(\tmods, \tkb) \models \tquery$ can be derived from some accepted word $w$.
\end{inparaenum}
This is formalized next, by the notion of TKB-modification \emph{abstraction} and the subsequent result.
\begin{definition}[TKB-modification abstraction]
		\label{def:abstraction}
    Consider a TKB $\tkb=(\Tmc, \Aseq)$, with $\Aseq = \Amc_0 \cdots \Amc_\ell$,
    a TCQ $\tquery$, and let $\DFA = (\alphabet, S, s_0, \gamma, F)$ 
    be the RT-DFA for $\tkb$ and $\tquery$.
    A word $w=w_0\cdots w_m\in\alphabet^*$, 
    inducing a finite run $\ppath=s_0\trans{w_0}\cdots\trans{w_{m}}s_{m+1}$
    in $\DFA$,
    is an \emph{abstraction} of 
    (or \emph{abstracts}) a TKB-modification
    $\tmods=\tmods_0\cdots\tmods_m$ iff, 
    for $j=0,\ldots, m$: 
    
        \begin{itemize}
            \item $w_j=\delsym$ and $\tmods_j = \del$; or
            
            \item
            $w_j = (\fixsym, \Upsilon)$ and, for $s_j=(Z,i)$,
            $\tmods_j=\fix{\mods}$, 
            with $(\Tmc, \apply(\mods, \Amc_i))\models \bigvee_{\Phi \in \Upsilon} \chformula(\Phi)\land
            \bigwedge_{\Phi \not\in \Upsilon} \lnot\chformula(\Phi)$; or
            
            \item 
            $w_j = (\addsym, \Upsilon)$ and
            $\tmods_j=\add{\mods}$, 
            with $(\Tmc, \apply(\mods, \emptyA))\models \bigvee_{\Phi \in \Upsilon} \chformula(\Phi)\land
            \bigwedge_{\Phi \not\in \Upsilon} \lnot\chformula(\Phi)$.
        \end{itemize}
        When this holds, $\tmods$ is an \emph{instantiation} of (or \emph{instantiates}) $w$.
\end{definition}

\begin{restatable}{theorem}{thmsoundcompl}\label{thm:rt-sound-compl}
	Consider a TKB $\tkb=(\Tmc,\Aseq)$, with $\Aseq=\Amc_0 \cdots \Amc_\ell$,
	a TCQ $\tquery$, and let $\DFA=(\alphabet,S,s_0,\gamma,F)$ 
	be the RT-DFA for $\tkb$ and $\tquery$.
	Then:
	\begin{enumerate}
		\item
			for every word $w \in \alphabet^*$, there exists an instantiation $\tmods$ s.t.~$w \in \L(\DFA)$ iff $\apply(\tmods,\tkb) \models \tquery$;
		\item
			for every word $w \in \alphabet^*$ and every instantiation $\tmods$ of $w$,  it holds that $w \in \L(\DFA)$ iff $\apply(\tmods,\tkb) \models \tquery$;
		\item
			for every TKB-modification $\tmods$ there exists a unique abstraction $w_\tmods$ s.t.~$w_\tmods \in \L(\DFA)$ iff $\apply(\tmods,\tkb) \models \tquery$.
	\end{enumerate}
\end{restatable}

Thm.~\ref{thm:rt-sound-compl} states that the language of $\DFA$ 
characterizes the set of solutions for the TKB-alignment of $\tkb$ against the 
specification $\tquery$; in particular, Item~2 ensures that every 
instantiation of some TKB-modification abstraction 
$w \in \L(\DFA)$ is a solution to TKB Alignment. 
Then, every optimal solution $\tmods^*$ is s.t.:

\begin{center}
	$\tmods^*=\argmin_\eta\set{\cost(\tmods,\tkb)\mid\tmods\text{ instantiates some } w\in\L(\DFA)}$.
\end{center}

Based on this, we can reduce the problem of finding $\tmods^*$ to that of finding a minimal path in a suitably weighted graph.

\begin{definition}[Minimal-instantiation Graph]\label{def:mmg}
	Consider a TKB $\tkb=(\Tmc,\Aseq)$, with $\Aseq = \Amc_0 \cdots \Amc_\ell$, a TCQ $\tquery$, and let $\DFA = (\alphabet, S, s_0, \gamma, F)$ be the RT-DFA for $\tkb$ and $\tquery$.
	The \emph{minimal-instantiation graph} for $\tkb$ and $\tquery$ is the weighted graph $\MMG = (N, E, \weight)$, where:
\begin{enumerate}
	\item
		$N = S$ is the finite set of nodes;
	\item
		$E \subseteq N \times \atmods \times N$, is the finite set of edges, labelled by atomic TKB-modifications;
	\item
		$\weight:E \ra \mathbb{R}^+$ is the edge weight function;
	\item\label{it:mmg-1}
		it holds that $e=((Z,i),\tmods,(Z', i'))\in E$ and $\weight(e)=c$ iff, for some $X$, $(Z',i')=\gamma((Z,i), X)$, and:
	\begin{itemize}[leftmargin=*, label=-]
		\item
			if $X = \delsym$ then $\tmods = \del$ and $c = \cost(\del,\Amc_i)$;
		\item
			if $X = (\fixsym, \Upsilon)$ then $\tmods = \\ \argmin_{\fix{\mods}} \bigset{\cost(\fix{\mods}, \Amc_i) \mid (\Tmc, \apply(\mods, \Amc_i)) \models \bigvee_{\Phi \in \Upsilon} \chformula(\Phi) \land \bigwedge_{\Phi \not\in \Upsilon} \lnot \chformula(\Phi)}$ and $c = \cost(\tmods, \Amc_i)$;
		\item
			if $X = (\addsym, \Upsilon)$ then $\tmods =\\ \argmin_{\add{\mods}} \bigset{\cost(\add{\mods}, \emptyAseq) \mid \\ (\Tmc, \apply(\mods, \emptyA)) \models \bigvee_{\Phi \in \Upsilon} \chformula(\Phi) \land \bigwedge_{\Phi \not\in \Upsilon} \lnot \chformula(\Phi)}$ and $c = \cost(\tmods, \emptyA)$.
	\end{itemize}
\end{enumerate}
\end{definition}
\begin{restatable}{lemma}{migcomplexity}\label{lem:mig-complexity}
The Minimal-instantiation Graph for a TKB  $\tkb$ and a TCQ $\tquery$ can be computed in triply exponential time and has triply exponential size wrt $\tquery$.
\end{restatable}

The minimal-instantiation graph $\MMG$ is a graph whose edges are labelled with atomic TKB-modifications
and weighted with the corresponding cost. Through its labels, every finite path $s_0\trans{\tmods_0}\cdots\trans{\tmods_{m-1}}s_m$ 
of $\MMG$ defines an instantiation $\tmods=\tmods_0\cdots\tmods_{m}$ of some input word (not necessarily accepted)
$w=w_0\cdots  w_{m}$ of $D$.
Also the viceversa holds, i.e., every input word $w$ of $\DFA$ is an abstraction of the TKB-modification $\tmods$ 
defined by some path of $G$. 

Observe that, by Item~\ref{it:mmg-1} of Def.~\ref{def:mmg}, $\tmods$ includes only minimal-cost atomic TKB-modifications $\tmods_i$,
thus it is a minimal-cost TKB-modification among all those that instantiate the same $w$.
Moreover, recall that, by Theorem~\ref{thm:rt-sound-compl}, every solution to TKB Alignment is associated to an abstraction $w \in \L(\DFA)$.
Thus, since every such $w$ has a minimal-cost instantiation in some path of $\MMG$, we can search for the minimal-cost
solution by exploring the paths of $\MMG$. Indeed, it is enough to search for the minimal-cost paths of $\MMG$ which 
correspond to the words $w$ accepted by $\DFA$;  since the nodes of $\MMG$ correspond to the states of $\DFA$, this corresponds
to searching for a minimal-path of $G$ starting in $s_0$ and ending in some node that is an accepting state for $\DFA$.


\begin{restatable}{theorem}{thmtkbalign}\label{thm:tkb-alignment}
	Consider a TKB $\tkb=(\Tmc,\Aseq)$, with $\Aseq = \Amc_0 \cdots \Amc_\ell$, a TCQ $\tquery$, and let $\DFA=(\alphabet,S,s_0,\gamma,F)$ 
	be the RT-DFA for $\tkb$ and $\tquery$.
	A TKB-modification $\tmods=\tmods_0\cdots\tmods_m$ is an optimal solution of TKB Alignment for $\tkb$ and $\tquery$ iff 
	there exists a minimum-cost path $\pi=n_0\trans{\tmods_0}\cdots\trans{\tmods_{m-1}}n_m$ in $\MMG$, s.t.~$n_0=s_0$
	and $n_m\in F$.
\end{restatable}
Thus, with $\MMG$ at hand, the search can easily be performed by, e.g., Dijkstra's algorithm.
However, in order to state the effective solvability of TKB Alignment, we must still guarantee that $\MMG$ is actually 
computable. In this respect, observe that we still need to explain how the labels and the weights of $\MMG$ can be obtained. 

By Def.~\ref{def:mmg} (Item~\ref{it:mmg-1}), 
computing the labels and the weights of $\MMG$ 
requires to solve, for every edge, one \emph{local minimization problem} of 
the form
$\argmin_{(\sigma,\mods)} \set{\cost((\sigma,\mods),\Aseq)}$, 
subject to a constraint of the form
$(\Tmc, \apply(\mods, \Amc)) \models \query$, with $\query\in \bcBCQ$.
This is the \emph{KB-alignment} problem, 
which we formally define and 
solve in Section~\ref{sec:kbalign}.
We report here that the problem can be solved in doubly exponential time.
This, together with the complexity results reported above, leads to the 
following.
\begin{restatable}{theorem}{tkbalgsolv}\label{thm:tkb-alg-solv}
TKB Alignment is solvable in triply exponential time.
\end{restatable}


As mentioned above, one might consider the semantics on TCQs given in~\cite{BaaderBL15} and instantiate TKB alignment with it.
Such variation on the semantics has minimal impact on the solution technique, which can be adapted to that setting with minimal changes. Details are omitted, due to space constraints.

\section{The KB-Alignment Problem}
\label{sec:kbalign}

In this section, we define the \emph{knowledge base alignment problem (KB Alignment)}, 
and show how to solve it for the query language $\bcBCQ$ and the DL $\alc$.
%

%

\begin{definition}[KB Alignment]
    \label{def:kb-alignment}
    Given a KB $\Kmc=\tup{\Tmc, \Amc}$ and  a query $\bquery \in \bcBCQ$, the 
    problem of KB Alignment consists in finding an ABox-modification $\mods$ 
    such that for $\Amc' = \apply(\mods,\Amc)$: 
    \begin{enumerate}
    	\item\label{prop:cons}
    	$\tup{\Tmc,\Amc'}$ is consistent;
    	\item\label{prop:entail}
    	$\tup{\Tmc,\Amc'}\models\bquery$; and
    	\item\label{prop:optimal}
    	$\cost(\mods) \leq \cost(\mods')$ for all ABox-modifications $\mods'$ satisfying Conditions~\ref{prop:cons} and \ref{prop:entail}. 
    \end{enumerate}
    An ABox-modification $\mods$ satisfying 
    Condition~\ref{prop:cons} and~\ref{prop:entail} is a 
    \emph{solution to KB Alignment} for $\tup{\Kmc,\bquery}$; if it also 
    satisfies~\ref{prop:optimal}, it is an \emph{optimal} solution.
\end{definition}
%
%
To solve an instance of KB Alignment, it is sufficient to consider ABox-modifications with operations defined using concept and role names from the input KB and query, i.e., using names from their signature. A \emph{signature} is a finite subset of $\cset \cup \rset$.  We use $\sig(X)$ to denote the set of concept and role names occurring in $X$, where $X$ can be a KB or a query.
\begin{restatable}{lemma}{sigrestr}\label{lem:signature:restriction}	
   Let $\Kmc$ be a KB and $\bquery \in \bcBCQ$. If KB Alignment has a solution for $\tup{\Kmc,\bquery}$, then it has an optimal solution $\mods$ where all ABox-operations $\ins{\assert}$ and $\rem{\assert}$ in $\mods$ are such that $\assert$ is an assertion defined over $\sig(\Kmc) \cup \sig(\bquery)$.
\end{restatable}
An algorithm that solves KB Alignment for $\alc$ KBs and queries in $\bcBCQ$ is introduced next.

\subsection{Solving KB-Alignment}\label{sub:sec:kb:alignment}

Algorithm~\ref{alg:kb:alignment} describes our method to solve KB Alignment. 
The approach is to  
(1) compute an upper bound on the cost of optimal 
solutions (if any), 
(2) iterate over all ABox-modifications of decreasing 
cost starting from the upper bound, and 
finally output a modification that satisfies Condition~\ref{prop:cons} to \ref{prop:optimal} from 
Definition~\ref{def:kb-alignment}, i.e., which is optimal.
 
More precisely, for the input  TBox $\Tmc$ and query $\bquery$, the algorithm first computes an ABox $\Amc'$ 
such that $\tup{\Tmc, \Amc'}$ is consistent and $\tup{\Tmc, \Amc'} \models \bquery$. 
Then, a  modification $\mods^*$ from the input ABox $\Amc$ into $\Amc'$ is 
obtained which always exists:
given two ABoxes $\Amc$ and $\Amc'$, a 
\emph{trivial modification from $\Amc$ into $\Amc'$} is defined as a 
sequence $\mods = \op_1\cdots \op_k \cdots \op_n$, where 
$\op_1 \cdots \op_k$ consists of the removal of all assertions in 
$\Amc$ and $\op_{k+1} \cdots \op_n$ consists of the insertion of all 
assertions occurring in $\Amc'$.
 %
  %
  %
  
  The algorithm computes one trivial modification, which, by 
  Definition~\ref{def:kb-alignment}, is a (possibly non-optimal) solution, thus $\cost(\mods^*)$ realizes the first step of the approach as it  
  is an upper bound on the cost of optimal  
  solutions. 
  Then, the \emph{for-loop} enumerates all ABox-modifications with cost smaller 
  than $\cost(\mods^*)$, that satisfy Condition~\ref{prop:cons} 
  and~\ref{prop:entail} of Definition~\ref{def:kb-alignment}, and returns one ABox modification of 
  minimal cost.

By using Lemma~\ref{lem:signature:restriction}, it is not hard to show that the 
output of Algorithm~\ref{alg:kb:alignment} is always an optimal solution.
 \begin{restatable}{lemma}{algcorrect}\label{lem:alg:1:sound:complete}
    If Algorithm~\ref{alg:kb:alignment} returns $\mods^*$ on input $\Kmc$ and 
    $\bquery$, then $\mods^*$ is an optimal solution to KB Alignment for 
    $\tup{\Kmc,\bquery}$.
 	If Algorithm~\ref{alg:kb:alignment} returns ``no solution'', 
 	then KB Alignment has no solution for $\tup{\Kmc,\bquery}$.
\end{restatable}

\begin{algorithm}[t]
	\caption{KB Alignment.}\label{alg:kb:alignment}
	\textbf{Input}: An $\alc$ KB $\Kmc=\tup{\Tmc, \Amc}$ and a query $\bquery \in \bcBCQ$.
	
	\textbf{Output}: An optimal solution of KB Alignment for $\tup{\Kmc,\bquery}$, if a solution exists; or ``no solution'', otherwise.
	
	\begin{algorithmic}[1]	
		\STATE Compute an ABox $\Amc'$ s.t.~$\tup{\Tmc, \Amc'}$ is consistent, $\tup{\Tmc, \Amc'} \models \bquery$ and $\sig(\Amc') \subseteq \sig(\Kmc) \cup \sig(\bquery)$. If  no such ABox exists, \textbf{return} ``no solution'';
		\label{alg1:line:nosolution}
		\STATE Define a trivial modification $\mods^*$ from $\Amc$ into $\Amc'$;\label{alg1:line:trivial}%
		\STATE Let $M$ be the set of ABox-modifications $\mods$ defined over $\sig(\Kmc) \cup \sig(\bquery)$ s.t.~$\cost(\mods) < \cost(\mods^*)$;  		 
		\FORALL{$\mods \in M$}
		\IF{($\mods$ satisfies conditions~\ref{prop:cons} and \ref{prop:entail} in Definition~\ref{def:kb-alignment}) \AND \\ \quad ($\cost(\mods) < \cost(\mods^*)$)  }\label{alg1:line:if}%
		\STATE \hspace*{0.5cm}
        $\mods^* := \mods$;
		\ENDIF
		\ENDFOR
		\RETURN{$\mods^*$};
	\end{algorithmic}
\end{algorithm}

Hence, to see that Algorithm~\ref{alg:kb:alignment} solves the KB-alignment 
problem, it remains to show that it terminates, i.e., all its steps can \emph{effectively} 
be computed. The following arguments 
show this for most of the 
steps:
\begin{itemize}[leftmargin=*] 		
	\item
	Since  $\Amc$ and $\Amc'$ are  finite sets, a trivial modification  $\mods^*$ can easily be computed from the ABox $\Amc'$.
	\item
	Consistency of $\alc$ KBs is a decidable problem (\cite{DL-Handbook-03}), as well as entailment of $\bcBCQ$-queries in $\alc$ (see Sec.~\ref{sub:sec:cqs}), hence the conditions at line~\ref{alg1:line:if} can be effectively verified for each $\mods \in M$.
	\item
	The set $M$ contains only ABox-modifications defined over the finite 
	signature $\sig(\Kmc) \cup \sig(\bquery)$. Hence, given $n > 0$, $M$ 
	contains finitely many $\mods$ with $n$ ABox-operations, and each such 
	$\mods$ has cost of at least $c\cdot n$, where $c$ is the minimal cost of 
	an ABox operation defined over $\sig(\Kmc) \cup \sig(\bquery)$.
	This implies that $M$ is a finite set and contains only modifications with 
	no more than $\cost(\mods^*) / c$ operations. Thus, $M$ can be visited in 
	finite time.
\end{itemize}
It remains to specify how to compute the initial ABox $\Amc'$ (or to determine that it does not exist).  This requires a more involved argument presented in the following.

\subsubsection{Computing the initial ABox $\Amc'$}

The computation of the initial ABox $\Amc'$ in our algorithm is closely related to the \emph{query emptiness problem} in ontology-mediated query answering.
This problem was introduced and investigated in \cite{BaaderBL16} for various DLs (including $\alc$) and the query language \CQ. We define this problem here for the more general query language $\bcCQ$.
It uses the notion of $\Sigma$-ABox, which refers to ABoxes that use only names from a signature $\Sigma$.

\begin{definition}[$\bcCQ$-query Emptiness]\label{def:query:emptiness}
	Let $\Tmc$ be a TBox, $\Sigma$ a signature and $\bquery \in \bcCQ$. 
	The query $\bquery$ is called \emph{empty for $\Sigma$ given $\Tmc$} if
	for all $\Sigma$-ABoxes $\Amc$ such that $(\Tmc,\Amc)$ is consistent, we have 
	$\cans{(\Tmc,\Amc)}{\bquery} = \emptyset$.

	\emph{$\bcCQ$-query Emptiness} is the problem of deciding, given a TBox 
	$\Tmc$, a signature $\Sigma$, and $\bquery \in \bcCQ$, whether $\bquery$ is 
	empty for $\Sigma$ \wrt $\Tmc$. 	
\end{definition}
 
 In \cite{BaaderBL16}, it is shown that to decide \CQ-query Emptiness in $\alc$, 
 it suffices to consider a single $\Sigma$-ABox $\Amc_{\Tmc,\Sigma}$. 
 This ABox is of exponential size and can be computed (from a given satisfiable TBox $\Tmc$ and a signature $\Sigma$) in exponential time, in the size of $\Tmc$ and the cardinality of $\Sigma$. Moreover, it satisfies the following:
 \begin{itemize}[leftmargin=*]
 	\item
 	the KB $(\Tmc,\Amc_{\Tmc,\Sigma})$ is consistent, and
 	\item
 	given a pure \cq\ $\query$, $\query$ is empty for $\Sigma$ \wrt $\Tmc$ iff \\ $\cans{(\Tmc,\Amc_{\Tmc,\Sigma})}{\query} = \emptyset$.
 \end{itemize}
  The arguments used to prove the second property can be easily extended to pure 
	${\bcCQ}$-queries, as the following result shows. 
  
  \begin{restatable}{lemma}{ASigmaenough}\label{lem:ASigmaT:is:enough:for:BbcCO}
  	Let $\Tmc$ be a satisfiable $\alc$ TBox, $\Sigma$ a signature and $\bquery \in \bcCQ$ a pure query. Then, $\bquery$ is empty for $\Sigma$ wrt\ $\Tmc$ iff $\cans{(\Tmc,\Amc_{\Tmc,\Sigma})}{\query} = \emptyset$.
\end{restatable}

For \emph{non-pure} queries, query emptiness can be reduced to the case of pure queries. This can be done as follows. 
Let $\Tmc$ be an $\alc$ TBox, $\Sigma$ a signature, and $\bquery \in \bcCQ$ a \emph{non-pure} query with $\ind(\bquery)=\{a_1,\ldots,a_m\}$. We select $m$ \emph{fresh} concept names $A_1,\ldots,A_m$, i.e., concept names neither occurring in $\Tmc, \bquery$ nor $\Sigma$. 
Then, we  define an $\alc$ TBox  $\Tmc_p$, a signature $\Sigma_p$ and a query $\bquery_p\in\bcCQ$, as follows:
\begin{itemize}
	\item
	$\Tmc_p = \Tmc \cup \Tmc_{\disj}$, where $\Tmc_{\disj} =\{A_i \sqcap A_j \sqsubseteq \bot \mid 1 \leq i < j \leq m\}$,
	\item
	$\bquery_p = \bquery_x \land \query_{\disj}$, where $\bquery_x$ is obtained from $\bquery$ by replacing each $a \in \ind(\bquery)$ by a fresh free variable $x_a$, whereas $\query_{\disj}$ is the \cq\ $\query_{\disj} = A_1(x_{a_1}) \land \ldots \land A_m(x_{a_m})$, and
	\item
	$\Sigma_p = \Sigma \cup \{A_1,\ldots,A_m\}$.
\end{itemize}
%
%
The following lemma shows that testing whether $\bquery$ is empty for $\Sigma$ \wrt $\Tmc$ reduces to checking emptiness of $\bquery_p$ for $\Sigma_p$ \wrt  $\Tmc_p$. Since $\bquery_p$ is a pure query, this yields a reduction from query emptiness of \emph{non-pure} queries to the case of pure queries.
%
 %
 \begin{restatable}{lemma}{removeindividuals}\label{lem:remove:individuals}
	%
	Let $\Tmc$ be an $\alc$ TBox, $\Sigma$ a signature, and $\bquery$ a query in $\bcCQ$. The following holds:
	\begin{enumerate}
		\item
		$\bquery$ is empty for $\Sigma$ \wrt $\Tmc$ iff $\bquery_p$ is empty for $\Sigma_p$ \wrt $\Tmc_p$.
		%
		%
		\item\label{point:2:lemma:nonpure:to:pure}
		%
		If there is a $\Sigma_p$-ABox $\Amc_p$ that witnesses non-emptiness of $\bquery_p$ for $\Sigma_p$ \wrt $\Tmc_p$, then, given $t \in \cans{(\Tmc_p,\Amc_p)}{\query_p}$, $\Amc_p$ can be transformed in polynomial time (in the size of $\Amc_p$ and $t$) into a $\Sigma$-ABox $\Amc$ witnessing non-emptiness of $\bquery$ for $\Sigma$ \wrt $\Tmc$. %
	\end{enumerate}
\end{restatable}

Hence, if the input query $\bquery$ of Algorithm~\ref{alg:kb:alignment} is pure, by Lemma~\ref{lem:ASigmaT:is:enough:for:BbcCO} the search space for the ABox $\Amc'$ can simply be restricted to $\{\Amc_{\Tmc,\Sigma}\}$ where $\Sigma=\sig(\Kmc)\cup\sig(\bquery)$.
%
%
Otherwise, Lemma~\ref{lem:remove:individuals} tells us how to obtain $\Amc'$ (if it exists). Namely, the algorithm first constructs $\Tmc_p, \bquery_p$ and $\Sigma_p$ from $\Tmc, \bquery$ and $\Sigma$. It then checks whether $\bquery_p$ is non-empty for $\Sigma_p$ \wrt $\Tmc_p$, by using $\Amc_{\Tmc_p,\Sigma_p}$.
If the latter is true, then  $\Amc'$ is selected as the ABox obtained from applying to $\Amc_{\Tmc_p,\Sigma_p}$ the transformation from the second statement in Lemma~\ref{lem:remove:individuals}. 
Overall, this provides a way to compute $\Amc'$ whenever it exists.

Hence, Algorithm~\ref{alg:kb:alignment} always terminates. This, together with Lemma~\ref{lem:alg:1:sound:complete}, yields solvability of KB Alignment. A closer look at Algorithm~\ref{alg:kb:alignment} reveals that it runs in \emph{double exponential time} in the size of the input KB and query. Thus, we obtain the following result.
%
\begin{restatable}{theorem}{solvabilitykbalignment}\label{th:solvability:kb:alignment:no:nominals:CQ}
	KB Alignment is solvable for $\alc$ and $\bcBCQ$ in \emph{double exponential time}.
\end{restatable}

\section{Discussion and Future Work}
\label{sec:con}

TKB Alignment is a new variant of the alignment problem that admits richer state and property descriptions. Our setting uses \alc-TKBs, CQs with \ltl operators, and a cost function for the edit operations. 
We have shown that TKB Alignment \wrt temporal CQs is solvable, by developing computation methods for both TKB and KB Alignment.

The TKB-alignment problem is closely related to abduction and to computing repairs of KBs, as these tasks also change a KB to either gain a desired consequence or remove an unwanted one. However, although being active research topics, neither of the two has yet been investigated for the temporalized setting and entailment of TCQs. Furthermore, TKB Alignment requires a cost-optimal solution, which is not very common in the context of abduction or repairs.

Interestingly, TKB Alignment can also be used for relaxing temporal CQ answering. Given a tuple of individuals $\bar{a}$ which is not a certain answer of a TCQ $\phi$, solve TKB Alignment for the Boolean TCQ obtained from $\phi'$ by assigning $\bar{a}$ to the answer variables of $\phi$. The costs computed during TKB Alignment for $\phi'$ then measure the \enquote{distance} to a certain answer of the query.

Our initial investigation on TKB Alignment uses a unitary cost measure for the edit operations mostly to ease presentation, as our approach can handle other cost measures easily. In this work, we did not regard rigid symbols, which are left for future work.

\clearpage

\ack
The work of Giuseppe Perelli was partially funded by MUR under the PRIN programme, grant B87G22000450001 (PINPOINT), and by the PNRR MUR project PE0000013-FAIR.
The work of Fabio Patrizi was partially funded by MUR under the PNRR MUR project PE0000013-FAIR, the ERC Advanced Grant WhiteMech
(No. 834228), and the Sapienza Project MARLeN.

\bibliography{ecai}

\iftrue 
\clearpage
\appendix
\section{Proofs}

\subsection{Proofs from Section~\ref{sec:tkb-align}}

\typing*

\begin{proof}	
	Consider a FO interpretation $\I$ and the type defined as $\Phi_{\I} = \set{\pquery \in \props(\ptquery) \mid \I \models \query}$.
	Clearly, we have $\I \models \chformula(\Phi_{I})$ and $\I \not\models \chformula(\Phi)$ for every other type $\Phi$.
	
	On the other hand, note that every type $\Phi \in 2^{\props(\ptquery)}$ identifies a combination of positive and negative boolean BCQs that are consistent by construction, and therefore admit a model $\I$ for all of them, which means that $\I \models \chformula(\Phi)$, and so that $\I$ is of type $\Phi$.
	
	The proof easily lifts to (infinite) sequences of FO traces and types.
	
\end{proof}

\thmbridge*
\begin{proof}	
	Observe that from Theorem~\ref{thm:sat-bstates}, we obtain that $\word \in \L(\PA_{\ptquery})$ iff $\word \models \ptquery$, regarded as propositional \ltl.
	Therefore, we need to prove that, for every $\seqI$, it holds that $\seqI \models \tquery$ iff $\word \models \ptquery$, with $\word$ being the trace type of $\seqI$.
	We do it by induction on the structure of $\tquery$.
	
	\begin{itemize}
		\item 
			Assume $\tquery = \psi \in \BCQ$ being a boolean conjunctive query.
			It holds that $\seqI \models \psi$ iff $\seqI_{0} \models \psi$, which, from Lemma~\ref{lem:typing}, is true iff $\hat{\psi} \in \word_0$, being $\word_0$ the type of $\seqI_{0}$.
			The latter holds iff $\word \models \hat{\psi}$ and so iff $\word \in \L(\PA_{\ptquery})$.
	\end{itemize}
	All the other (boolean and temporal) cases are standard proofs that build on top of the semantics of \tcq and of propositional \ltl.
\end{proof}

\reductcomplexity*
\begin{proof}
Direct consequence of Thm.~\ref{thm:sat-bstates}, which states that the state space of $\PA_\tquery$ has doubly exponential size in $\tquery$, and the EXPTIME-completeness of deciding satisfiability of $\chi(\Phi)$ wrt $\Tmc$~\cite{Lutz08}.
\end{proof}

\lemtail*
\begin{proof}	
	We prove the two implications separately.
	First, assume that $q_k \notin \Acc(\PA^\Tmc_{\ptquery})$ and let $\word^k = \Phi_{k} \Phi_{k + 1} \cdots$ be a type trace and $\seqI^{k} = \I_{k} \I_{k + 1}$ a corresponding FO-trace.
	If $\I_i \not\models \Tmc$ for some $i \geq k$ then the trace type $\word$ already satisfies the statement.
	Assume instead that $\I_i \models \Tmc$ fore every $i \geq k$ and  consider the run
	$$\ppath^{\Tmc} = q_k \trans{\Phi_{k}} q_{k + 1} \trans{\Phi_{k + 1}} \cdots$$
	of $\word^{k}$ in $\PA^{\Tmc}_{\ptquery}$.
	First, observe that, since $\I_i \models \Tmc$ for every $i \geq k$, we have that $\chformula(\Phi_i)$ is consistent with $\Tmc$ for every $i \geq k$.
	This means that the extra state $q^{*}$ in $\PA^{\Tmc}_{\ptquery}$ does not occur in $\ppath^{\Tmc}$, which implies that $\ppath^{\Tmc}$ is also the run of $\word^{k}$ in $\PA_{\ptquery}$.
	By concatenation, we obtain that
	$$\ppath' = q_0 \trans{\Phi_0} \cdots \trans{\Phi_{k - 1}} q_k \trans{\Phi_{k}} q_{k + 1} \trans{\Phi_{k + 1}} \cdots$$
	is the run in $\PA_{\ptquery}$ of the word $\word$.
	Now, since $q_k \notin \Acc(\PA^\Tmc_{\ptquery})$, we obtain that $\min\set{\painf(\ppath^{\Tmc}, \PA^\Tmc_{\ptquery})}$ is odd.
	%
	%
	Therefore, by the definition of $\col^\Tmc$, we obtain that 
	$\min\set{\painf(\ppath', \PA_{\ptquery})} = \min\set{\painf(\ppath^{\Tmc}, \PA_{\ptquery})} = \min\set{\painf(\ppath^{\Tmc}, \PA^{\Tmc}_{\ptquery})} - 1$ is even and then that $\word \in \L(\PA_{\ptquery})$.
	By Theorem~\ref{thm:bridge} we derive that the FO trace $\seqI$, being of type $\word$, is such that $\seqI \models \tquery$.
	
	We now prove the other direction by counter-nominal argument.
	Assume $q_k \in \Acc(\PA^{\Tmc}_{\ptquery})$.
	This means that there exists a word $\word^k = \Phi_{k} \Phi_{k + 1} \ldots$ whose run $$\ppath^{\Tmc} = q_k \trans{\Phi_{k}} q_{k + 1} \trans{\Phi_{k + 1}} \cdots$$ of $\word^{k}$ in $\PA^{\Tmc}_{\ptquery}$ starting from $q_k$ is accepting.
	This implies that $q^{*}$ never appears in $\ppath^{\Tmc}$, which means that $\chformula(\Phi_i)$ is consistent with $\Tmc$, for every $i \geq k$, and that $$\ppath' = q_0 \trans{\Phi_0} \cdots \trans{\Phi_{k - 1}}q_k \trans{\Phi_{k}} q_{k + 1} \trans{\Phi_{k + 1}} \cdots$$ is the run in $\PA_{\ptquery}$ of the word $\word = \Phi_0 \cdots \Phi_{k - 1} \Phi_{k} \Phi_{k + 1} \cdots$.
	Being that $\ppath^{\Tmc}$ is accepting in $\PA^{\Tmc}_{\ptquery}$, we have that $\min\set{\painf(\ppath^{\Tmc}, \PA^\Tmc_{\ptquery})}$ is even, which implies that $\min\set{\painf(\ppath', \PA_{\ptquery})} = \min\set{\painf(\ppath^{\Tmc}, \PA_{\ptquery})} = \min\set{\painf(\ppath^{\Tmc}, \PA^{\Tmc}_{\ptquery})} - 1$ is odd and then that $\word \notin \L(\PA_{\ptquery})$.
	By Theorem~\ref{thm:bridge}, every FO trace $\seqI$ of type $\word$ is such that $\seqI \not\models \tquery$.
	Observe that, being $\chformula(\Phi_i)$ consistent with $\Tmc$ for every $i \geq k$, this means that $\I_i \models \Tmc$ for every $i \geq k$.
	This proves the statement.
	
\end{proof}

\rtdacomplexity*
\begin{proof}
Let $\tkb=(\Tmc,\Aseq)$ and $\PA_{\ptquery}^{\Tmc}=(2^{\props(\ptquery)},Q^\Tmc,\delta^{\Tmc},q_0,\col^{\Tmc})$ 
	be the $\Tmc$-reduct of $\PA_{\ptquery}=(2^{\props(\ptquery)},Q,\delta,q_0,\col)$. 
 Define the RT-DFA for $\tkb$ and $\tquery$ as $\DFA=(\alphabet,S,s_0,\gamma,F)$.

The result follows directly from the definition of RT-DFA (Def.~\ref{def:rtdfa})
and the following facts: the sizes of $Q$ and \alphabet are doubly exponential wrt \tquery, which implies a triply exponential size of $S$; checking whether there exists an ABox $\Amc$ s.t.~$(\Amc,\Tmc)\models \phi$ is time-exponential wrt both $\Tmc$ (this is shown in the proof of Thm~\ref{th:solvability:kb:alignment:no:nominals:CQ}) and $\phi$, with $\phi=\bigvee_{\Phi \in \Upsilon}\chformula(\Phi)\land
	                        			\bigwedge_{\Phi \not\in \Upsilon}\lnot\chformula(\Phi)$ 
doubly exponential wrt $\tquery$; and computing $\Acc(\PA_{\ptquery}^{\Tmc})$ is doubly exponential wrt \tquery.
\end{proof}

In order to prove Thm.~\ref{thm:rt-sound-compl}, the following auxiliary result is needed.
\begin{lemma}\label{lem:abstraction-instantiation}
	For every word $w \in \alphabet^{*}$ there exists a TKB-modification $\tmods$ that instantiates $w$.
	Moreover, for every TKB-modification $\mods$ there exists a unique abstraction $w$ of it.
\end{lemma}

\begin{proof}
	We first prove that for every word $w$ there exists an instantiation $\tmods$ of it.
	Consider also the run $\ppath = s_0 \trans{w_0} \ldots \trans{w_{m}} s_{m + 1}$ be the corresponding run on the RT-DFA.
	Then, for each $j \leq m$, we have:
	
	\begin{itemize}
		\item 
			If $w_j = \delsym$ then define $\tmods_j = \del$;
			
		\item 
			if $w_j = (\fixsym, \Upsilon)$ and $s_j = (Z, i)$, then consider an ABox $\Amc$ consistent with $\Tmc$ such that $(\Tmc, \Amc) \models \bigvee_{\Phi \in \Upsilon} \chformula(\Phi)\land
			\bigwedge_{\Phi \not\in \Upsilon} \lnot\chformula(\Phi)$.
			Observe that such ABox always exists.
			Now, let $\mods$ be a KB-modification such that $\Amc = \apply(\mods, \Amc_i)$.
			Also such $\mods$ always exists.
			Then, define $\tmods_{j} = \fix{\mods}$;
			
		\item 
			if $w_j = (\addsym, \Upsilon)$ and $s_j = (Z, i)$, then consider an ABox $\Amc$ consistent with $\Tmc$ such that $(\Tmc, \Amc) \models \bigvee_{\Phi \in \Upsilon} \chformula(\Phi)\land
			\bigwedge_{\Phi \not\in \Upsilon} \lnot\chformula(\Phi)$.
			Observe that such ABox always exists.
			Now, let $\mods$ be a KB-modification such that $\Amc = \apply(\mods, \emptyA)$.
			Also such $\mods$ always exists.
			Then, define $\tmods_{j} = \add{\mods}$.
	\end{itemize}
	
	Clearly, the TKB-modification $\tmods$ satisfies Definition~\ref{def:abstraction} and so it is an instantiation of $w$.
	
	For the second statement, consider a TKB modification $\tmods$ of length $m$.
	We construct by induction on $m$ the unique abstraction of $\tmods$.
	
	As base case, if $m = 0$, we have three possible cases:
	
	\begin{enumerate}
		\item 
			If $\tmods_0 = \delsym$ then define $w_{0} = \delsym$;
			
		\item 
			If $\tmods_0 = \fix{\mods}$ then consider $\Amc = \apply(\mods, \Amc_0)$ and $\Upsilon$ being the \emph{unique} element in $2^{2^{\props}}$ such that $(\Tmc, \Amc) \models \bigvee_{\Phi \in \Upsilon} \chformula(\Phi)\land \bigwedge_{\Phi \not\in \Upsilon} \lnot\chformula(\Phi)$.
			Therefore, define $w_0 = (\fixsym, \Upsilon)$;
			
		\item 
			If $\tmods_0 = \add{\mods}$ then consider $\Amc = \apply(\mods, \emptyA)$ and $\Upsilon$ being the \emph{unique} element in $2^{2^{\props}}$ such that $(\Tmc, \Amc) \models \bigvee_{\Phi \in \Upsilon} \chformula(\Phi)\land \bigwedge_{\Phi \not\in \Upsilon} \lnot\chformula(\Phi)$.
			Therefore, define $w_0 = (\addsym, \Upsilon)$.
	\end{enumerate}

	Clearly, in all the cases, $w_0$ is the only abstraction of $\tmods_0$.
	
	For the induction case, assume there is a unique abstraction for every TKB modification of length $m$ and consider the TKB-modification $\tmods$ of length $m + 1$.
	First, consider $\tmods' = \tmods_0 \ldots \tmods_m$ be the prefix of $\tmods$ up to $m$.
	By induction hypothesis, let $w'$ be the only abstraction of $\tmods'$ and $\ppath = s_0 \trans{w_0} \ldots \trans{w_{m}} s_{m + 1}$ be the corresponding run on the RT-DFA, with $s_m = (Z, i)$.
	We distinguish three cases:
	
	\begin{enumerate}
		\item 
			If $\tmods_{m + 1} = \del$, then define $w_{m + 1} = \delsym$;
			
		\item 
			If $\tmods_{m + 1} = \fix{\mods}$, then consider $\Amc = \apply(\mods, \Amc_i)$ and $\Upsilon$ being the \emph{unique} element in $2^{2^{\props}}$ such that $(\Tmc, \Amc) \models \bigvee_{\Phi \in \Upsilon} \chformula(\Phi)\land \bigwedge_{\Phi \not\in \Upsilon} \lnot\chformula(\Phi)$.
			Therefore, define $w_{m + 1} = (\fixsym, \Upsilon)$;
			
		\item 
			If $\tmods_{m + 1} = \add{\mods}$, then consider $\Amc = \apply(\mods, \emptyA)$ and $\Upsilon$ being the \emph{unique} element in $2^{2^{\props}}$ such that $(\Tmc, \Amc) \models \bigvee_{\Phi \in \Upsilon} \chformula(\Phi)\land \bigwedge_{\Phi \not\in \Upsilon} \lnot\chformula(\Phi)$.
			Therefore, define $w_{m + 1} = (\addsym, \Upsilon)$;
	\end{enumerate}

	Clearly, also in this case, the abstraction $w = w' \cdot w_{m + 1}$ is the unique abstraction of $\tmods$.
	
\end{proof}

\thmsoundcompl*

\begin{proof}
	First, observe that thanks to Lemma~\ref{lem:abstraction-instantiation} we need to prove only Item~2 of the theorem.
	Indeed, Item~1 would follow from Item~2 and from the fact that every abstraction admits at least an instantiation.
	Analogously, Item~3 would follow from Item~2 and from the fact that every TKB-modification admits a unique abstraction.
	
	We then prove Item~2.
	For technical convenience, we prove a slightly more general result.
	First, for every TKB $\tkb = (\Tmc, \Aseq)$ with $\Aseq = \Amc_0 \cdots \Amc_\ell$ and an index $i$, define $\tkb_{\geq i} = (\Tmc, \Aseq_{\geq i})$ where $\Aseq_{\geq i} = \Amc_i \cdots \Amc_{\ell}$.
	We prove that, for every word $w \in \alphabet^{*}$ and state $s = (Z, i)$ of the automaton $\DFA$, it holds that the run 
	
	\[
	\ppath = s \trans{w_0} s_1 \trans{w_1} \ldots \trans{w_m} s_{m + 1}
	\]
	
	is accepting in $\DFA$ iff, for all instantiations $\tmods$ of $w$ and every trace type $\word$ such that  $\seqI \models \apply(\tmods, \tkb_{\geq i})$ for every $\seqI$ of type $\word$, it holds that the run in $\PA_{\ptquery}$ starting from some $q \in Z$ over $\word$ is accepting.
	
	Item~2 of the theorem then follows by setting $s = (\{q_0, 0\})$ the initial state of $\DFA$, which in turns applies to the TKB $\tkb$.
	
	The proof is by induction on the length $m$ of $w$ and, subsequently, $\tmods$.
	
	As base case, assume $m = 0$, therefore both $m$ and $\tmods$ are the empty sequence.
	This implies that the path $\ppath = s$ is made by just $s = (Z, i)$ and it is accepting iff $s \in F$.
	By construction of $\DFA$, first we have that $i = \ell + 1$ and so that $\tkb_{\geq i} = (\Tmc, \emptyAseq)$.
	Moreover, we have that $Z \cap \Acc(\PA^{\Tmc}_{\ptquery}) = \emptyset$ and so that every state $q$ in $Z$ does not belong to $\Acc(\PA^{\Tmc}_{\ptquery})$.
	By Lemma~\ref{lem:tail}, we have that every trace type $\word$ such that $\seqI \models (\Tmc, \emptyAseq)$ for every $\seqI$ of type $\word$, is such that the run starting from $q$ is accepting in $\PA_{\ptquery}$, which proves the statement.
	
	For the induction case, assume that the property holds for a given $m$, and that $w$ and $\tmods$ are of length $m + 1$.
	Observe that, since $\ppath$ is accepting, then also the path
	
	\[
	\ppath' = s_1 \trans{w_1} \ldots \trans{w_m} s_{m + 1}
	\]
	is accepting, which means that, by induction hypothesis, for every instantiation $\tmods'$ of $w' = w_{1} \cdots w_{m}$, every trace type $\word'$ such that  $\seqI \models \apply(\tmods, \tkb_{\geq 1})$ for every $\seqI'$ of type $\word'$, it holds that the run in $\PA_{\ptquery}$ starting from some $q \in Z_1$, with $s_1 = (Z_1, \iota)$ over $\word'$ is accepting.
	We distinguish three cases:
	
	\begin{itemize}
		\item 
			$w_0 = \delsym$, then $\tmods_{0} = \del$ and every instantiation of $w$ is of the form $\tmods_{0} \cdot \tmods'$.
			We also obtain that $s_1 = \gamma((Z, 0), \delsym) = (Z, 1)$.
			In addition, observe that 
			
			\[
			\apply(\tmods_{0} \cdot \tmods', \tkb_{\geq 0}) = \apply(\del \cdot \tmods', \tkb_{\geq 0}) = \apply(\tmods', \tkb_{\geq 1})
			\]
			which, combined with the induction hypothesis, proves the statement.
			
		\item 
			$w_0 = (\fixsym, \Upsilon)$ and so $\tmods_{0} = \fix{\mods}$ for some $\mods$ such that $(\Tmc, \apply(\mods, \Amc_0)) \models \bigvee_{\Phi \in \Upsilon}\chformula(\Phi)\land
			\bigwedge_{\Phi \not\in \Upsilon}\lnot\chformula(\Phi)$.
			Also, every instantiation of $w$ is of the form $\tmods_{0} \cdot \tmods'$.
			Observe that 
			
			\[
			s_1 = \gamma((Z,0), (\fixsym, \Upsilon)) = (Z_1, 1)
			\]
			
			where $Z_1$ contains only states $q'$ such that $\delta(q,\Phi)=q'$, for some $q \in Z$.
			
			Observe that			
			
			\begin{align*}
				\apply(\tmods_{0} \cdot \tmods', \tkb_{\geq 0}) & =\apply(\fix{\mods} \cdot \tmods', \tkb_{\geq 0}) \\ 
				& =(\Tmc, \apply(\mods, \Amc_0) \cdot \apply(\tmods', \Aseq_{\geq 1}))
			\end{align*}
		
			By induction hypothesis, every trace type $\word'$ such that $\seqI' \models (\Tmc, \apply(\tmods', \Aseq_{\geq 1}))$ for every $\seqI'$ of type $\word'$ has an accepting run in $\PA_{\ptquery}$ starting from $s_1$.
			Moreover, the only types $\Phi_0$ entailed by $(\Tmc, \apply(\mods, \Amc_0))$ must belong to $\Upsilon$ by construction.
			Therefore, the word $\word = \Phi_0 \cdot \word'$ clearly is accepted by $\PA_{\ptquery}$ from some state $q \in Z$ such that $\delta(q, \Phi_0) \in Z_1$ and so the statement holds.
			
			\item 
				$w_0 = (\addsym, \Upsilon)$ and so $\tmods_{0} = \add{\mods}$ for some $\mods$ such that $(\Tmc, \apply(\mods, \emptyA)) \models \bigvee_{\Phi \in \Upsilon}\chformula(\Phi)\land
				\bigwedge_{\Phi \not\in \Upsilon}\lnot\chformula(\Phi)$.
				Also, every instantiation of $w$ is of the form $\tmods_{0} \cdot \tmods'$.
				Observe that 
				
				\[
				s_1 = \gamma((Z,0), (\fixsym, \Upsilon)) = (Z_1, 0)
				\]
				where $Z_1$ contains only states $q'$ such that $\delta(q,\Phi)=q'$, for some $q \in Z$.
				Observe that
				
				\begin{align*}
					\apply(\tmods_{0} \cdot \tmods', \tkb_{\geq 0}) & =\apply(\add{\mods} \cdot \tmods', \tkb_{\geq 0}) \\ 
					& =(\Tmc, \apply(\mods, \emptyA) \cdot \apply(\tmods', \Aseq_{\geq 0}))
				\end{align*}
				By induction hypothesis, every trace type $\word'$ such that $\seqI' \models (\Tmc, \apply(\tmods', \Aseq_{\geq 0}))$ for every $\seqI'$ of type $\word'$ has an accepting run in $\PA_{\ptquery}$ starting from $s_1$.
				Moreover, the only types $\Phi_0$ entailed by $(\Tmc, \apply(\mods, \emptyA))$ must belong to $\Upsilon$ by construction.
				Therefore, the word $\word = \Phi_0 \cdot \word'$ clearly is accepted by $\PA_{\ptquery}$ from some state $q \in Z$ such that $\delta(q, \Phi_0) \in Z_1$ and so the statement holds.
	
	\end{itemize}

	Note that all the arguments in both base and induction cases are equivalences, which means both directions of Item~2 are proved.
\end{proof}

\migcomplexity*
\begin{proof}
Consequence of Def.~\ref{def:mmg}, Lemma~\ref{lem:rtda-complexity}, which states that RT-DFA $\DFA = (\alphabet, S, s_0, \gamma, F)$ for $\tkb$ and $\tquery$ has triply eponential size wrt $\tquery$ and the fact that KB-Alignment is solvable in doubly exponential time (see Thm.~\ref{th:solvability:kb:alignment:no:nominals:CQ}).
\end{proof}

\tkbalgsolv*
\begin{proof}
    Immediate consequence of the fact that the Minimal-instantiation Graph $G$ has triply exponential size and that Shortest Path can be solved in polynomial time on $G$.
\end{proof}



\subsection{Proofs from Section~\ref{sec:kbalign}}

\sigrestr*
\begin{proof}
	Let $\Kmc = (\Tmc,\Amc)$, $\mods=\op_0\cdots \op_n$  be an optimal solution to KB Alignment for $(\Kmc,\bquery)$ and let $\Amc' = \apply(\mods,\Amc)$. 

	Suppose that $\mods$ contains an ABox operation $\op_i=\rem{\alpha}$ such that $\alpha \not \in \Amc$. In addition, assume that there is no $j, 0 \leq j < i$ such that $\op_j=\op_i$. We make the following case distinction:
	\begin{itemize}
		\item
		$\ins{\alpha}$ does not occur in $\op_0\cdots \op_{i-1}$. Then, removing $\op_i$ from $\mods$ yields an ABox-modification $\mods'$ such that $\Amc' = \apply(\mods',\Amc)$ and $\cost(\mods') < \cost(\mods)$.
		\item
		There is $j, 0 \leq j < i$ such that $m_j=\ins{\alpha}$. In this case, removing $\op_i$ and all ABox operations $m_\ell = \ins{\alpha}$ ($0 \leq \ell < i$) from $\mods$ 
		results in a shorter ABox-modification of lesser cost than $\mu$ and that would also result in $\Amc'$ if applied to $\Amc$.
	\end{itemize}
	Hence, we can assume  without loss of generality that each ABox operation $\rem{\alpha}$ occurring in $\mods$ is such that $\alpha \in \Amc$.
	
	Consider the ABox-modification $\mods^*$ obtained from $\mods$ by removing all operations of the form $\ins{\alpha}$ such that $\alpha$ is not defined over $\sig(\Kmc) \cup \sig(\bquery)$. Two observations regarding $\mods^*$ are in order:
	\begin{itemize}
		\item
		Since every ABox operation $\rem{\alpha}$ in $\mods$ satisfies that $\alpha \in \Amc$ (by assumption), the same is the case for $\mods^*$. Hence, every ABox operation in $\mods^*$ is defined over $\sig(\Kmc) \cup \sig(\bquery)$.
		\item
		By definition of $\mods^*$, we have that $\cost(\mods^*) \leq \cost(\mods')$. In addition, $\Amc^* \subseteq \Amc'$ where $\Amc^* = \apply(\mods^*,\Amc)$.
		Hence, since $(\Tmc,\Amc')$ is consistent, this means that $(\Tmc,\Amc^*)$ is consistent as well.
	\end{itemize}	

	Thus, to conclude the proof of the lemma, it remains to show that $(\Tmc,\Amc^*) \models \bquery$. Let $\Imc$ be a model of $(\Tmc,\Amc^*)$. Since no individual name occurs in $\Tmc$ and assertions removed from $\Amc'$ are defined over concept and role names not contained in $\sig(\Tmc) \cup \sig(\bquery)$, it is not hard to see that: 
	\begin{itemize}
		\item
		$\Imc$ can be extended into a model $\Jmc$ of $(\Tmc,\Amc')$ such that $\Imc$ and $\Jmc$ are identical on $\sig(\Tmc) \cup \sig(\bquery)$ and on the interpretation of the individuals in $\ind(\bquery)$.
		\item
		 Let $\query'$ be a \CQ  occurring in $\bquery$. Since $\Imc$ and $\Jmc$ are identical on $\sig(\query')$ and $\ind(\query')$, it follows that:
		\begin{equation}\label{eq:base:case:signature:lemma}
			\J \models \query'\ \text{iff}\ \ \I \models \query'.
		\end{equation}
		Hence, by using induction on the structure of $\bquery$ one can easily show that $\J \models \bquery$ iff $\I \models \bquery$. The base case follows from \eqref{eq:base:case:signature:lemma}, since it corresponds to the \cq s occurring in $\bquery$.
		Consequently, since $(\Tmc,\Amc') \models \bquery$, it follows that $\Jmc \models \bquery$ and $\Imc\models\bquery$.						
	\end{itemize}
	
	Finally, since $\Imc$ is an arbitrary model of $(\Tmc,\Amc^*)$, we have thus shown that  $(\Tmc,\Amc^*) \models \bquery$. This completes the proof.	
\end{proof}

\algcorrect*
\begin{proof}
	Suppose that Algorithm~\ref{alg:kb:alignment} returns an ABox-modification $\mods^*$ on input $\Kmc$ and $\bquery$. The following observations imply that $\mods^*$ is an optimal solution to KB Alignment for $(\Kmc,\bquery)$.
	\begin{itemize}
		\item
		The modification $\mods^*$ is either the trivial modification computed at Line~\ref{alg1:line:trivial}, or a modification in $M$ satisfying the test at Line~\ref{alg1:line:if}. Hence, $\mods^*$ is a solution to KB Alignment for $(\Kmc,\bquery)$, because it satisfies Condition~\ref{prop:cons} and \ref{prop:entail} in Definition~\ref{def:kb-alignment}.
		\item
		Since $\mods^*$ is a solution to KB Alignment for $(\Kmc,\bquery)$, it follows that some optimal solution $\mods'$ to KB Alignment for $(\Kmc,\bquery)$ exists. 
		By Lemma~\ref{lem:signature:restriction}, we can assume that $\mods'$ is defined over $\sig(\Kmc) \cup \sig(\bquery)$. In addition, since $\mods^*$ is a solution, it follows that 
		\begin{equation}\label{eq:fact}
	   	   \cost(\mods') \leq \cost(\mods^*).
		\end{equation}
	If $\cost(\mods') = \cost(\mods^*)$, then $\mods^*$ is obviously optimal. Otherwise, inequation \eqref{eq:fact} implies that $\cost(\mods') < \cost(\mods^*)$. This in turn implies that $\mods'\in M$, since either $\mods^*$ is the trivial modification computed at Line~\ref{alg1:line:trivial} in Algorithm~\ref{alg:kb:alignment} or $\mods^* \in M$.
	Obviously, the \emph{for-loop} would never choose $\mods^*$ if $M$ contains a modification with smaller cost. Hence, $\cost(\mods') < \cost(\mods^*)$ cannot be the case. Therefore, $\mods^*$ must be an optimal solution.
	\end{itemize}

	Suppose that Algorithm~\ref{alg:kb:alignment} returns \enquote{no solution}. Due to Line~\ref{alg1:line:nosolution} there is then no ABox $\Amc'$ such that:
	\begin{itemize}
		\item
		$(\Tmc,\Amc')$ is consistent, 
		\item $(\Tmc,\Amc') \models \bquery$, and 
		\item $\sig(\Amc') \subseteq \sig(\Kmc) \cup \sig(\bquery)$.
	\end{itemize}  
     The non-existence of such an ABox $\Amc'$ implies that there is no solution $\mods$ to KB Alignment for $(\Kmc,\bquery)$ such that $\mods$ is defined over $\sig(\Kmc) \cup \sig(\bquery)$. Note that any such ABox-modification $\mods$ would yield an ABox defined over the signature $\sig(\Kmc) \cup \sig(\bquery)$, like $\Amc'$ is.
     Thus, the application of Lemma~\ref{lem:signature:restriction} yields that KB Alignment has indeed no solution for $(\Kmc,\bquery)$.
\end{proof}

\ASigmaenough*

To prove this lemma, we slightly extend the proof arguments given in \cite{BaaderBL16} for the \CQ query language to the more general query  language \bcCQ. To this end, we need to introduce some notions and auxiliary results from \cite{BaaderBL16}.
Let us start with the notion of an ABox homomorphism.
\begin{definition}
	Let $\Amc$ and $\Amc'$ be two ABoxes. An \emph{ABox homomorphism} from $\Amc$ to $\Amc'$ is a total function $h:\ind(\Amc) \mapsto \ind(\Amc')$ satisfying the following conditions:
	\begin{itemize}
		\item
		$A(a) \in \Amc$ implies $A(h(a)) \in \Amc'$, and
		\item 
		$r(a,a') \in \Amc$ implies $r(h(a),h(a')) \in \Amc'$.
	\end{itemize}
\end{definition}
 The following lemma states an important property of ABox homomorphisms for query answering. It generalizes Lemma~11 in \cite{BaaderBL16} from \CQ to \bcCQ.
\begin{lemma}\label{lem:queries:canonical:abox}
	Let $\Tmc$ be an $\alc$ TBox, $\bquery\in \bcCQ$ with free variables $x_1,\ldots,x_k$, $\Amc$ and $\Amc'$  ABoxes, and $\hphism$ an ABox homomorphism from $\Amc$ to $\Amc'$. If $(\Tmc,\Amc) \models \bquery[(a_1,\ldots,a_k)]$ then $(\Tmc,\Amc') \models \bquery[(\hphism(a_1),\ldots,\hphism(a_k))]$.
\end{lemma}
\begin{proof}
	We assume that $(\Tmc,\Amc) \models \bquery[(a_1,\ldots,a_k)]$ holds.	
	The case where $(\Tmc,\Amc')$ is not consistent is trivial. 
	If $(\Tmc,\Amc')$ is consistent, it has a model. Let $\Imc$ be an arbitrary model of $(\Tmc,\Amc')$. We show that 
	\begin{equation}\label{eq:hom:preserves:entailment}
		\Imc \models \bquery[(\hphism(a_1),\ldots,\hphism(a_k))].
	\end{equation}
	To prove this, we define an interpretation $\Jmc$ from $\Imc$ by reinterpreting the individual names in $\Amc$, i.e
	\begin{itemize}
		\item
		$\Delta^\Jmc := \Delta^\Imc$,
		\item
		$X^\Jmc := X^\Imc$ for all $X \in \cset\cup\rset$, and
		\item
		$a^\Jmc := \hphism(a)^\Imc$ for all $a \in \ind(\Amc)$.
	\end{itemize}
	Let $\query_a$ be a \cq\ occurring in $\bquery[(a_1,\ldots,a_k)]$, and $\bquery_\hphism$ be the \cq\ that results from replacing all occurrences of all $a_i$ ($1 \leq i \leq k$)
	in $\query_a$ by $\hphism(a_i)$. Hence, 	since ${a_i}^\Jmc = \hphism(a_i)^\Imc$ for all $i, 1 \leq i \leq k$, the following holds:
	\begin{equation}\label{eq:base:case}
		\Jmc \models \query_a\ \ \ \text{ iff }\ \ \ \Imc \models \query_\hphism
	\end{equation}
	Therefore, by using induction on the structure of $\bquery[(a_1,\ldots,a_k)]$ and applying \eqref{eq:base:case} to the base case, one can easily show that:
	\begin{equation*}
		\Jmc \models \bquery[(a_1,\ldots,a_k)]\ \text{ iff }\ \Imc \models \bquery[(\hphism(a_1),\ldots,\hphism(a_k))]
	\end{equation*}
	Hence, since $(\Tmc,\Amc) \models \bquery[(a_1,\ldots,a_k)]$, to  show that \eqref{eq:hom:preserves:entailment} holds it is enough to show that $\Jmc \models (\Tmc,\Amc)$.
	\begin{itemize}
		\item
		Since no individual name occurs in $\Tmc$, $\Imc \models \Tmc$ implies that $\Jmc$ is also a model of $\Tmc$. 
		\item
		To see that $\Jmc$ is a model of $\Amc$ as well, consider any concept assertion $A(a) \in \Amc$. The homomorphism $\hphism$ ensures that $A(\hphism(a)) \in \Amc'$. Hence, $\I \models \Amc'$ implies that $\hphism(a)^\I \in A^\I$, and it follows by definition of $\J$ that $a^\J \in A^\J$. The case of role assertions $r(a,b) \in \A$ can be shown analogously.
	\end{itemize}
	Overall, we have just shown that any model $\Imc$ of $(\Tmc,\Amc')$ satisfies \eqref{eq:hom:preserves:entailment}. Thus, $(\Tmc,\Amc') \models \bquery[(\hphism(a_1),\ldots,\hphism(a_k))]$.
\end{proof}

The ABox $\Amc_{\Tmc,\Sigma}$ defined in \cite{BaaderBL16} is \emph{canonical} in the following sense. For any satisfiable $\alc$ TBox $\Tmc$, the following holds:
\begin{enumerate}
	\item
	 $(\Tmc,\Amc_{\Tmc,\Sigma})$ is consistent. 
	 \item\label{claim:BBL16-2}
	 A $\Sigma$-ABox $\Amc$ satisfies that $(\Tmc,\Amc)$ is consistent iff there is an ABox homomorphism from $\Amc$ to $\Amc_{\Tmc,\Sigma}$. 
\end{enumerate}
Using these two properties about $\Amc_{\Tmc,\Sigma}$ and Lemma~\ref{lem:queries:canonical:abox}, we are now ready to conclude the proof of Lemma~\ref{lem:ASigmaT:is:enough:for:BbcCO}.
 
 \medskip
 
 \begin{proof}[Proof of Lemma~\ref{lem:ASigmaT:is:enough:for:BbcCO}]
 	The \emph{left-to-right} direction is trivial. For the opposite direction, assume that $\bquery$ is non-empty for $\Sigma$ \wrt $\Tmc$. Then, there exists a $\Sigma$-ABox $\Amc$ such that $\cans{(\Tmc,\Amc)}{\bquery} \neq \emptyset$.
 	Thus, since there exists an ABox homomorphism from $\Amc$ to $\Amc_{\Tmc,\Sigma}$ (by \ref{claim:BBL16-2}.\ above), Lemma~\ref{lem:queries:canonical:abox} yields that $\cans{(\Tmc,\Amc_{\Tmc,\Sigma})}{\bquery} \neq \emptyset$.
 \end{proof}

\removeindividuals*
%

\begin{proof}
	Let $x_1,\ldots,x_k$ be the free variables of $\bquery$ and $a_1,\ldots,a_m$ the individual names occurring in $\bquery$. Then, $\bquery_x$ has $k+m$ free variables, i.e.
	\begin{equation*}
		\fvarsq(\bquery_x) =\{x_1,\ldots,x_k\} \cup \bigcup\limits_{i=1}^{m} \{x_{a_i}\},
	\end{equation*}
	where each $x_{a_i}$ is fresh variable not occurring in $\bquery$. We represent the free variables as the ordered tuple $(x_1,\ldots,x_k, x_{a_1},\ldots,x_{a_m})$.
	
	Let us start with Claim 1, by showing the \emph{left-to-right} implication.
	
	($\Rightarrow$) Assume that $\bquery_p$ is non-empty for $\Sigma_p$ \wrt $\Tmc_p$. Hence, there exists a $\Sigma_p$-ABox $\Amc_p$ such that  $(\Tmc_p,\Amc_p)$ is consistent and  $\cans{(\Tmc_p,\Amc_p)}{\bquery_p} \neq \emptyset$. Thus, there exists  a tuple $t=(b_1,\ldots,b_{k+m})$ of individual names of $\iset$ such that $t \in \cans{(\Tmc_p,\Amc_p)}{\bquery_p}$. 
	The following assumptions about $\Amc_p$ are without loss of generality:
	\begin{enumerate}[(a)]
		\item
		 $\Amc_p$ contains no individuals from $\ind(\bquery)$. This can be assumed because neither $\bquery_p$ nor $\Tmc_p$ contain individual names,		
		\item
		$b_{k+1},\ldots,b_{k+m}$ are all distinct individuals of $\Amc_p$. 
		This follows from the fact that $(\Tmc_p,\Amc_p)$ is consistent, the disjointness axioms occurring $\Tmc_{\disj}$ and the form of  $\bquery_{\disj}$.
		\item
		For all $i \in \{1,\ldots,m\}$, $\Amc_p$  contains no assertion of the form $A_i(b)$ with $b \neq b_{k+i}$. Such an assertion would either contradict $t \in \cans{(\Tmc_p,\Amc_p)}{\bquery_p}$, or would not be necessary to ensure that $t$ belongs to $\cans{(\Tmc_p,\Amc_p)}{\bquery_p}$.
	\end{enumerate}
	Let us define a $\Sigma$-ABox $\Amc$ from $\Amc_p$, by taking the following steps.
	\begin{enumerate}
		\item
		Remove all assertions of the form $A_i(b)$ occurring in $\Amc_p$ ($1 \leq i \leq m$).
		\item
		Rename every remaining individual name $b_{k+i}$ as $a_i$ ($1 \leq  i \leq m$).
	\end{enumerate}
     This renaming is well-defined since all individual names $b_{k+1},\ldots,b_{k+m}$  are distinct (recall assumption (b)). For all $i \in \{1,\ldots,k\}$, we write $\overline{b_i}$ to denote the following individual from $\{b_1,\ldots,b_k,a_1,\ldots,a_m\}$:
     \begin{itemize}
     	\item
     	$b_i$, if $b_i \neq b_{k+j}$  for all  $j \in \{1,\ldots,m\}$, or
     	\item
     	$a_j$, if $b_i = b_{k+j}$ for some  $j \in \{1,\ldots,m\}$.
     \end{itemize}
     %
	
	We show that $\Amc$ witnesses non-emptiness of $\bquery$ for $\Sigma$ \wrt $\Tmc$. 
	Clearly, by removing all assertions of the form $A_i(b)$ from $\Amc_p$, we have that $\Amc$ is a $\Sigma$-ABox. To see that $(\Tmc,\Amc)$ is consistent, consider a model $\Imc$ of $(\Tmc _p,\Amc_p)$. We extend $\Imc$ to interpret the individual names $a_i$ as:
	\begin{itemize}
		\item
		$a_i^\Imc = b_{k+i}^\Imc$ ($1 \leq i \leq m$).
	\end{itemize} 
     Since $\Imc \models \Amc_p$, by construction of $\Amc$ it is easy to see that $\Imc \models \Amc$ as well. Further, since no individual name occurs in $\Tmc$, the extended $\Imc$ is still a model of $\Tmc$. Hence, $(\Tmc,\Amc)$ is consistent.
	
	It remains to show that $\cans{(\Tmc,\Amc)}{\bquery} \neq \emptyset$. To this end, we show that $\bar{t} = (\overline{b_1},\ldots,\overline{b_k}) \in \cans{(\Tmc,\Amc)}{\bquery}$. 
	Let $\Jmc$ be an arbitrary model of $(\Tmc,\Amc)$. It is enough to prove that 
	\begin{equation}\label{eq:ent:to:show}
		\Jmc \models \bquery[\:\overline{b_1},\ldots,\overline{b_k}\:].
	\end{equation}
    We use $\Jmc$ to build an interpretation $\Jmc_p$ such that $\Jmc_p \models (\Tmc_p, \Amc_p)$. Since $t \in \cans{(\Tmc_p,\Amc_p)}{\bquery_p}$,  we would have that $\Jmc_p \models \bquery_p[t]$. Recall that $\bquery_p = \bquery_x \land \bquery_d$, $t=(b_1,\ldots,b_{k+m})$,  and $\bquery_x$ has $k+m$ free variables.
    Hence, we would have that:
    \begin{equation}\label{eq:ent:that:holds}
    	\Jmc_p \models \bquery_x[b_1,\ldots,b_k,b_{k+1},\ldots,b_{k+m}].
    \end{equation}
    This will then be used to show that \eqref{eq:ent:to:show} holds.
    
    Let $\Delta^{\Jmc}_0, \Delta^{\Jmc}_1 \ldots, \Delta^{\Jmc}_m$ be disjoint copies of $\Delta^\Jmc$. Given $d \in \Delta^\Jmc$, we denote by $d_i$ the corresponding copy in $\Delta^{\Jmc}_i$ ($0 \leq i \leq m$). The interpretation $\Jmc_p$ is defined as follows:
    \begin{itemize}
    	\item
    	$\Delta^{\Jmc_p} := \Delta^{\Jmc}_0 \cup \cdots \cup \Delta^{\Jmc}_m$,
    	\item
    	$B^{\Jmc_p} := \{d_i \in \Delta^{\Jmc}_i \mid d \in B^\Jmc \text{ and } 0 \leq i \leq m\}$\\[.2em] \hspace*{4.5cm} for all  $B \in \cset \setminus \Sigma_p$,
    	\item
    	$r^{\Jmc_p} := \{(d_i,e_j) \in \Delta^{\Jmc}_i \times \Delta^{\Jmc}_j \mid (d,e) \in r^\Jmc \text{ and }$ \\[.2em] 
    	\hspace*{3cm} $0 \leq i,j \leq m\}$, for all $r\in\rset$,
    	\item
    	$A_i^{\Jmc_p} := \{d_i \in \Delta^{\Jmc}_i \mid d =a_i^\Jmc\}$ for all $1 \leq i \leq m$. 
    	     %
    	
    	 %
    \end{itemize}
   The individual names $b_1,\ldots,b_{k+m}$  are interpreted as follows. 
   \begin{itemize}
   	   \item
   	   For all $i \in \{1,\ldots,k\}$ such that $\overline{b_i}\neq a_j$ ($1 \leq j \leq m$):
   	   \begin{equation*}
   	   	 b_i^{\Jmc_p} = d_0 \in \Delta_0^{\Jmc_p}, \text{ where } d=b_i^\Jmc.
   	   \end{equation*}

   	   \item
   	   For all $j \in \{1,\ldots,m\}$:
   	   \begin{equation*}
   	   	 b_{k+j}^{\Jmc_p} = d_j \in \Delta_j^{\Jmc_p}, \text{ where } d = a_{j}^{\Jmc}.
   	   \end{equation*}
   \end{itemize}
  This assignment covers all individuals names in $b_1,\ldots,b_{k+m}$, because if  $\overline{b_i}= a_j$ for some $i \in \{1,\ldots,k\}$ then $b_i=b_{k+j}$ for some $j \in \{1,\ldots,m\}$.
    
  To show that $\Jmc_p \models (\Tmc_p,\Amc_p)$, we use the property that $\alc$ is bisimulation invariant \cite{BaaderHLS17}.

  We start by defining a binary relation $\bis \subseteq \Delta^\Jmc \times \Delta^{\Jmc_p}$ as follows:
	\begin{equation*}
		\bis := \{(d,d_i) \in \Delta^\Jmc \times \Delta^{\Jmc_p} \mid 0 \leq i \leq m\}.
	\end{equation*}
It is not hard to show that  $\bis$ is a \emph{bisimulation} between $\Jmc$ and $\Jmc_p$ \wrt the symbols in $(\cset \cup \rset) \setminus \Sigma_p$. 
%
Hence, the bisimulation invariance of $\alc$ guarantees the following for all $\alc$ concepts $C$ defined over $(\cset \cup \rset) \setminus \Sigma_p$:
\begin{equation}\label{bisimulation:prop}
	d \in C^\Jmc\ \text{ iff }\ d_i \in C^{\Jmc_p},\ \text{for all } i \in \{0,\ldots,m\}.
\end{equation}
Therefore, since no concept name in $\Sigma_p$ occurs in $\Tmc$ and $\Jmc \models \Tmc$, the correspondence in \eqref{bisimulation:prop} implies that $\Jmc_p \models \Tmc$. In addition, the interpretation of $A_1,\ldots,A_m$ implies that $\Jmc_p \models \Tmc_d$. Thus, we have that $\Jmc_p \models \Tmc_p$.
	Regarding $\Amc_p$, we consider the possible forms of its assertions:
	\begin{itemize}
		\item
		$A_i(b) \in \Amc_p$ for some $i \in \{1,\ldots,m\}$. By assumption (c), we know that $b=b_{k+i}$. The definition of $\Jmc_p$ tells us that $b_{k+i}^{\Jmc_p}=d_i$ where $d=a_i^\Jmc$, and that $A_i^{\Jmc_p}=\{d_i\}$. Thus, $b_{k+i}^{\Jmc_p} \in A_i^{\Jmc_p}$.
		\item
		$A(b) \in \Amc_p$ for $A \in \Sigma$. If $b=b_{k+i}$ for some $i \in \{1,\ldots,m\}$, then $A(a_i) \in \Amc$. Hence, $\Jmc \models \Amc$ implies that $a_i^\Jmc \in A^\Jmc$.
		Let $a_i^\Jmc=d \in \Delta^\Jmc$. By construction of $\Jmc_p$, we have that $d_i \in A^{\Jmc_p}$ and $b_{k+i}^{\Jmc_p} = d_i$. Thus, it follows that $b_{k+i}^{\Jmc_p} \in A^{\Jmc_p}$.

		In case $b\neq b_{k+i}$, we have that $A(b) \in \Amc$. Hence, $b^\Jmc\in A^\Jmc$. It is easy to see from the definition of $\Jmc_p$ that $b^{\Jmc_p} \in A^{\Jmc_p}$.
		\item
		The case of the role inclusions in $\Amc_p$ can be shown similarly, by considering the interpretation of role names in $\Jmc_p$ and the renaming of $b_{k+i}$ into $a_i$.
	\end{itemize}

  Overall, we have shown that $\Jmc_p \models (\Tmc_p,\Amc_p)$. Hence, we have that \eqref{eq:ent:that:holds} holds. From this, we can show that \eqref{eq:ent:to:show} holds in two steps, as follows:
  \begin{enumerate}
  	\item
  	  We show that for each \cq\ $\query^*$ in $\bquery[\:\overline{t}\:]$ and the corresponding one $\query^*_x$ in $\bquery_x[t]$ it holds that:
  	  \begin{equation}\label{base:case:induction}
  	  	   \Jmc \models \bquery^*\ \ \text{ iff }\ \ \Jmc_p \models \bquery^*_x.
  	  \end{equation}
      This is a consequence of the definition of $\Jmc_p$ and the fact that $\bquery[\: \overline{t} \:]$ and $\bquery_x [t]$ are Boolean queries that are identical modulo renaming of $a_i$ and $b_{k+i}$.
      \item
     We apply induction on the structure of $\bquery[\: \overline{t} \:]$ to show that \eqref{eq:ent:that:holds} implies \eqref{eq:ent:to:show}, by using \eqref{base:case:induction} as the induction base case.
  \end{enumerate}

  This concludes the proof of the \emph{left-to-right} direction of Claim 1 of the lemma.
  %
    %
	We continue by showing the \emph{right-to-left} direction.
	
	($\Leftarrow$)
	Suppose that $\bquery$ is non-empty for $\Sigma$ \wrt $\Tmc$. This means that there is a $\Sigma$-ABox $\Amc$ and a tuple $t=(b_1,\ldots,b_k)$ of individual names $b_1,\ldots,b_k \in \iset$ such that $(\Tmc,\Amc)$ is consistent and $t \in\cans{(\Tmc,\Amc)}{\bquery}$.
	Since $\Tmc$ contains no occurrences of individual names, there exists a model $\Imc$ of $(\Tmc,\Amc)$ such that $a_i^\Imc \neq a_j^\Imc$ for all $i \neq j, 1 \leq i, j \leq m$. 
	
   Consider the ABox $\Amc_p = \Amc \cup \{A_i(a_i) \mid 1 \leq i \leq m\}$ and the interpretation $\Imc_p$ that extends $\Imc$ by defining $A_i^{\Imc_p} := \{a_i^\Imc\}$ for all $i \in \{1,\ldots m\}$. It is clear that $\Amc_p$ is a $\Sigma_p$-ABox. In addition, it is easy to see that $\Imc_p \models (\Tmc_p, \Amc_p)$. Hence, $(\Tmc_p,\Amc_p)$ is a consistent KB.
    Further, since $\Tmc \subseteq \Tmc_p$ and $\Amc \subseteq \Amc_p$, every model $\Jmc$ of $(\Tmc_p, \Amc_p)$ is a model of $(\Tmc,\Amc)$. This implies that $\Jmc \models \bquery[t]$. Since each individual $a_i$ is replaced by $x_{a_i}$ to obtain $\bquery_x$ from $\bquery$, it follows that $\Jmc \models \bquery_x[(b_1,\ldots,b_k,a_1,\ldots,a_m)]$.
    Moreover, the additional assertions in $\Amc_p$ ensure that $\Jmc \models \query_{\disj}[(a_1,\ldots,a_m)]$. Hence, we have that:
    \begin{equation*}
    	(b_1,\ldots,b_k,a_1,\ldots,a_m) \in \cans{(\Tmc_p,\Amc_p)}{\bquery_x \land \query_{\disj}}.
    \end{equation*} 
    
     Thus, we have shown that $\bquery_p$ is non-empty for $\Sigma_p$ \wrt $\Tmc_p$.\\
     
     To conclude, let us look at Claim 2 of the lemma. Suppose there exists a $\Sigma_p$-ABox $\Amc_p$ such that $(\Tmc_p,\Amc_p)$ is consistent and $\cans{(\Tmc_p,\Amc_p)}{\bquery_p} \neq \emptyset$. In addition, let $t \in \cans{(\Tmc_p,\Amc_p)}{\bquery_p}$.
     In the \emph{left-to-right} direction of the proof of Claim 1, we show how to transform $\Amc_p$ into a $\Sigma$-ABox $\Amc$ that witnesses non-emptiness of $\bquery$ for $\Sigma$ \wrt $\Tmc$.
     In particular, $\Amc$ is obtained by performing two operations:
     \begin{enumerate}
     	\item
     	Removing from $\Amc_p$ all assertions of the form $A_i(b)$ for some individual $b$ and $i \in \{1,\ldots,m\}$, and 
     	\item
     	Renaming all remaining occurrences of individual names $b_{k+i}$ ($1 \leq i\leq m$). 
     \end{enumerate}  
     This can be done by iterating over all assertions occurring in $\Amc_p$ and checking for the occurrence of $A_i$ and $b_{k+i}$. Since $t$ is a tuple of $k+m$ elements and all individuals names $b_{k+i}$ occur in $t$, this iteration takes polynomial time in the size of $\Amc_p$ and $t$.
\end{proof}

\solvabilitykbalignment*

\begin{proof}
	To see that KB Alignment is solvable in \emph{double exponential time}, we analyze the complexity of Algorithm~\ref{alg:kb:alignment}. In the following, we use $\mathcal{N}$ to denote the combined size of the KB $\Kmc$ and the query $\bquery$ in the input of Algorithm~\ref{alg:kb:alignment}.
	Let us continue by analyzing each step of the algorithm.
	\begin{itemize}				
		\item
		As explained in the paper, the initial ABox $\Amc'$  can be selected as $\Amc_{\Tmc,\Sigma}$ when the input query $\bquery$ is pure, where $\Sigma=\sig(\Kmc)\cup\sig(\bquery)$. Then, testing whether $(\Tmc,\Amc')$ entails $\bquery$ can be done in \emph{double exponential time} in $\mathcal{N}$. The reason is that $\Amc_{\Tmc,\Sigma}$  can be of exponential size, and entailment for $\bcBCQ$ in $\alc$ is an ExpTime-complete problem.

		In case $\bquery$ is \emph{non-pure}, the algorithm first checks whether $\bquery_p$ is non-empty for $\Sigma_p$ \wrt $\Tmc_p$. This can be done by looking directly at $\Amc_{\Tmc_p,\Sigma_p}$ and checking whether $\cans{(\Tmc_p,\Amc_{\Tmc_p,\Sigma_p})}{\bquery_p} \neq \emptyset$. The latter can be done in \emph{double exponential time} in $\mathcal{N}$: 
		\begin{itemize}
			\item
			Since the increase of the size of $\Tmc_p$ and $\Sigma_p$ \wrt $\Tmc$ and $\Sigma$ is polynomial in the size of the input query $\bquery$, it follows that  the ABox $\Amc_{\Tmc_p,\Sigma_p}$ is of size (at most) exponential and can be computed in exponential time in $\mathcal{N}$. 
			Therefore, checking whether $\cans{(\Tmc_p,\Amc_{\Tmc_p,\Sigma_p})}{\bquery_p} \neq \emptyset$ can be done in \emph{double exponential time} in $\mathcal{N}$.
		\end{itemize}
		
		If the  check $\cans{(\Tmc_p,\Amc_{\Tmc_p,\Sigma_p})}{\bquery_p}\neq \emptyset$ is positive, then there is $t\in \cans{(\Tmc_p,\Amc_{\Tmc_p,\Sigma_p})}{\bquery_p}$, which can be obtained while doing the check. Then, $\Amc'$ can be selected as the ABox that results from applying to $\Amc_{\Tmc_p,\Sigma_p}$ the transformation described in the proof of Lemma~\ref{lem:remove:individuals} \wrt the tuple $t$.
		This transformation is polynomial in the size of $\Amc_{\Tmc_p,\Sigma_p}$ and $t$. Since the arity of $t$ is lineal in the size of the input query $\bquery$, it follows that the obtained ABox  is of size at most exponential in $\mathcal{N}$.

		 Overall, we have shown that the initial ABox $\Amc'$ is of size at most exponential in $\mathcal{N}$, and that the first step of Algorithm~\ref{alg:kb:alignment} can be executed in \emph{double exponential time}.
		\item
		By definition of a trivial modification, the initial modification $\mods^*$ in Algorithm~\ref{alg:kb:alignment} is a sequence of the form 
		\begin{equation*}
			\op_1\cdots \op_k \cdots \op_n,
		\end{equation*}
	where $\op_1 \cdots \op_k$ consists of the removal of all assertions in $\Amc$ and $\op_{k+1} \cdots \op_n$ consists of the insertion of all assertions occurring in $\Amc'$. Hence, $\mods^*$ can be computed in exponential time in $\mathcal{N}$. Further, the magnitude of the number $\cost(\mods^*)$  is at most exponential in $\mathcal{N}$. 
	\item
	Regarding the set $M$, it satisfies the following (as analyzed in Subsetion~\ref{sub:sec:kb:alignment}):
	\begin{itemize}
		\item
		Each ABox-modication $\mods$ in $M$ is defined over the finite signature $\sig(\Kmc) \cup \sig(\bquery)$ and satisfies $\cost(\mods) < \cost(\mods^*)$, and
		\item
		$M$ contains only modifications with no more than $\cost(\mods^*) / c$ ABox-operations, where $c$ is the minimal cost of 
		an ABox operation defined over $\sig(\Kmc) \cup \sig(\bquery)$.		
	\end{itemize}
     Since the magnitude of $\cost(\mods^*)$ is at most exponential in $\mathcal{N}$, each modification $\mods \in M$ contains a number of ABox-operations that is at most exponential in $\mathcal{N}$. Moreover, since each ABox-operation is defined over a symbol in $\sig(\Kmc) \cup \sig(\bquery)$, $M$ can be enumerated in \emph{double exponential time} in the size of $\mathcal{N}$.
     \item
     In the enumeration of $M$, checking whether the conditional at Line~\ref{alg1:line:if} holds, amounts to:
     \begin{itemize}
     	\item
     	applying $\mods$ to the input ABox $\Amc$ to obtain an ABox $\Amc^*$,
     	\item 
     	checking whether $(\Tmc,\Amc^*)$ is consistent and $(\Tmc,\Amc^*) \models \bquery$, and
     	\item
     	verifying whether $\cost(\mods) < \cost(\mods^*)$, where $\mods^*$ is the modification of minimal cost encountered so far along the enumeration.
     \end{itemize}
       From the discussions in the previous points, $\Amc^*$ is of size at most exponential and can be computed in exponential time in $\mathcal{N}$.
       Moreover, since knowledge base consistency and entailment of $\bcBCQ$ can be checked in exponential time for $\alc$, then checking whether $(\Tmc,\Amc^*)$ is consistent and $(\Tmc,\Amc^*) \models \bquery$ can be done in \emph{double exponential time} in $\mathcal{N}$. Finally, since the magnitude of $\cost(\mods)$ and $\cost(\mods^*)$ are not greater than the initially computed upper bound, checking whether $\cost(\mods) < \cost(\mods^*)$ holds is not more costly than \emph{double exponential time}.
 	\end{itemize}
    
     Therefore, taking into account the previous analysis, we can conclude that Algorithm~\ref{alg:kb:alignment} runs in \emph{double exponential time} in $\mathcal{N}$. Thus, KB Alignment can be solved in \emph{double exponential time}.
\end{proof}

\fi
\end{document}